\documentclass[showpacs,preprintnumbers,amsmath,amssymb,nofootinbib]{revtex4}
\usepackage{latexsym,epsf}
\usepackage{graphicx}
\usepackage{bm}
\usepackage{psfig,epsfig}
\begin{document}
\preprint{{\vbox{\hbox{NCU-HEP-k014}
\hbox{Feb 2004}\hbox{rev. Aug 2004}
}}}

\def\ap#1#2#3{           {Ann. Phys. (NY) }{\bf #1} (19#2) #3}
\def\arnps#1#2#3{        {Ann. Rev. Nucl. Part. Sci. }{\bf #1} (19#2) #3}
\def\cnpp#1#2#3{        {Comm. Nucl. Part. Phys. }{\bf #1} (19#2) #3}
\def\apj#1#2#3{          {Astrophys. J. }{\bf #1} (19#2) #3}
\def\asr#1#2#3{          {Astrophys. Space Rev. }{\bf #1} (19#2) #3}
\def\ass#1#2#3{          {Astrophys. Space Sci. }{\bf #1} (19#2) #3}

\def\apjl#1#2#3{         {Astrophys. J. Lett. }{\bf #1} (19#2) #3}
\def\ass#1#2#3{          {Astrophys. Space Sci. }{\bf #1} (19#2) #3}
\def\jel#1#2#3{         {Journal Europhys. Lett. }{\bf #1} #2 (19#3)}

\def\ib#1#2#3{           {\it ibid. }{\bf #1} #2 (19#3)}
\def\nat#1#2#3{          {Nature }{\bf #1} #2 (19#3)}
\def\nps#1#2#3{          {Nucl. Phys. B (Proc. Suppl.) } {\bf #1} #2 (19#3)} 
\def\np#1#2#3{           {Nucl. Phys. }{\bf #1} #2 (19#3)}
\def\npp#1#2#3{           {Nucl. Phys. }{\bf #1} #2 (20#3)}
\def\pl#1#2#3{           {Phys. Lett. }{\bf #1} #2 (19#3)}
\def\pll#1#2#3{           {Phys. Lett. }{\bf #1} #2 (20#3)}
\def\pr#1#2#3{           {Phys. Rev. }{\bf #1} #2 (19#3)}
\def\prr#1#2#3{           {Phys. Rev. }{\bf #1} #2 (20#3)}
\def\prep#1#2#3{         {Phys. Rep. }{\bf #1} (19#2) #3}
\def\prl#1#2#3{          {Phys. Rev. Lett. }{\bf #1} #2 (19#3) }
\def\prll#1#2#3{          {Phys. Rev. Lett. }{\bf #1} #2 (20#3)}

\def\pw#1#2#3{          {Particle World }{\bf #1} (19#2) #3}
\def\ptp#1#2#3{          {Prog. Theor. Phys. }{\bf #1} #2 (19#3)}
\def\jppnp#1#2#3{         {J. Prog. Part. Nucl. Phys. }{\bf #1} (19#2) #3}

\def\rpp#1#2#3{         {Rep. on Prog. in Phys. }{\bf #1} (19#2) #3}
\def\ptps#1#2#3{         {Prog. Theor. Phys. Suppl. }{\bf #1} (19#2) #3}
\def\rmp#1#2#3{          {Rev. Mod. Phys. }{\bf #1} (19#2) #3}
\def\zp#1#2#3{           {Zeit. fur Physik }{\bf #1} (19#2) #3}
\def\fp#1#2#3{           {Fortschr. Phys. }{\bf #1} (19#2) #3}
\def\Zp#1#2#3{           {Z. Physik }{\bf #1} (19#2) #3}
\def\Sci#1#2#3{          {Science }{\bf #1} (19#2) #3}

\def\n.c.#1#2#3{         {Nuovo Cim. }{\bf #1} (19#2) #3}
\def\r.n.c.#1#2#3{       {Riv. del Nuovo Cim. }{\bf #1} (19#2) #3}
\def\sjnp#1#2#3{         {Sov. J. Nucl. Phys. }{\bf #1} (19#2) #3}
\def\yf#1#2#3{           {Yad. Fiz. }{\bf #1} (19#2) #3}
\def\zetf#1#2#3{         {Z. Eksp. Teor. Fiz. }{\bf #1} (19#2) #3}
\def\zetfpr#1#2#3{         {Z. Eksp. Teor. Fiz. Pisma. Red. }{\bf #1} (19#2) #3}
\def\jetp#1#2#3{         {JETP }{\bf #1} (19#2) #3}
\def\mpl#1#2#3{          {Mod. Phys. Lett. }{\bf #1} (19#2) #3}
\def\ufn#1#2#3{          {Usp. Fiz. Naut. }{\bf #1} (19#2) #3}
\def\sp#1#2#3{           {Sov. Phys.-Usp.}{\bf #1} (19#2) #3}
\def\ppnp#1#2#3{           {Prog. Part. Nucl. Phys. }{\bf #1} (19#2) #3}
\def\cnpp#1#2#3{           {Comm. Nucl. Part. Phys. }{\bf #1} (19#2) #3}
\def\ijmp#1#2#3{           {Int. J. Mod. Phys. }{\bf #1} (19#2) #3}
\def\ic#1#2#3{           {Investigaci\'on y Ciencia }{\bf #1} (19#2) #3}
\def\tp{these proceedings}
\def\pc{private communication}
\def\ip{in preparation}
\relax
\newcommand{\TeV}{\,{\rm TeV}}
\newcommand{\GeV}{\,{\rm GeV}}
\newcommand{\MeV}{\,{\rm MeV}}
\newcommand{\keV}{\,{\rm keV}}
\newcommand{\eV}{\,{\rm eV}}
\newcommand{\Tr}{{\rm Tr}\!}
\renewcommand{\arraystretch}{1.2}
\newcommand{\be}{\begin{equation}}
\newcommand{\ee}{\end{equation}}
\newcommand{\bea}{\begin{eqnarray}}
\newcommand{\eea}{\end{eqnarray}}
\newcommand{\ba}{\begin{array}}
\newcommand{\ea}{\end{array}}
\newcommand{\bc}{\begin{center}}
\newcommand{\ec}{\end{center}}
\newcommand{\bmat}{\left(\ba}
\newcommand{\emat}{\ea\right)}
\newcommand{\refs}[1]{(\ref{#1})}
\newcommand{\ler}{\stackrel{\scriptstyle <}{\scriptstyle\sim}}
\newcommand{\ger}{\stackrel{\scriptstyle >}{\scriptstyle\sim}}
\newcommand{\lag}{\langle}
\newcommand{\rag}{\rangle}
\newcommand{\ns}{\normalsize}
\newcommand{\cm}{{\cal M}}
\newcommand{\gr}{m_{3/2}}
\newcommand{\p}{\partial}
\newcommand{\bsg}{$b\rightarrow s + \g$}
\newcommand{\Bsg}{$B\rightarrow X_s + \g$}
\newcommand{\atal}{{\it et al.}}
\newcommand{\cq}{{\cal Q}}
\newcommand{\cqt}{{\widetilde {\cal Q}}}
\newcommand{\wtlc}{{\widetilde C}}
\def\321{$SU(3)\times SU(2)\times U(1)$}
\def\tl{{\tilde{l}}}
\def\tL{{\tilde{L}}}
\def\bd{{\overline{d}}}
\def\tL{{\tilde{L}}}
\def\a{\alpha}
\def\b{\beta}
\def\bsg{$ b \rightarrow s + \g$}
\def\g{\gamma}
\def\c{\chi}
\def\d{\delta}
\def\D{\Delta}
\def\db{{\overline{\delta}}}
\def\Db{{\overline{\Delta}}}
\def\e{\epsilon}
\def\f{\frac}
\def\tn{-\frac{2}{9}}
\def\tt{\frac{2}{3}}
\def\l{\lambda}
\def\n{\nu}
\def\m{\mu}
\def\nt{{\tilde{\nu}}}
\def\p{\phi}
\def\P{\Phi}
\def\k{\kappa}
\def\x{\xi}
\def\r{\rho}
\def\s{\sigma}
\def\t{\tau}
\def\th{\theta}
\def\ne{\nu_e}
\def\nm{\nu_{\mu}}
\def\snui{\tilde{\nu_i}}
\def\la{{\makebox{\tiny{\bf loop}}}}
\def\ti{\tilde}
\def\ssc{\scriptscriptstyle}
\def\wtl{\widetilde}
\def\mp{\marginpar}
\def\und{\underline}
\renewcommand{\Huge}{\Large}
\renewcommand{\LARGE}{\Large}
\renewcommand{\Large}{\large}

\title{ Radiative B decays in Supersymmetry without $R$-parity}

\author{ Otto C. W. Kong and Rishikesh D. Vaidya \\
{\ns\it  Department of Physics, National Central University,}\\
{\ns\it Chung-Li 32054 Taiwan.}}

\vskip 0.8 cm

\begin{abstract} 
We present a systematic analysis of all the contributions at the 
leading log order to the branching ratio 
of the inclusive radiative decay \Bsg\ 
in the framework of supersymmetry without $R$-parity. The relevant set of 
four-quark operators involved in QCD running  are extended from
6 (within SM and MSSM) to 24, with also many new contributions to the
Wilson coefficients of (chromo)magnetic penguins for either chiral structure.
We present complete analytical results here without any {\it a priori}
assumptions on the form of $R$-parity violation. Mass eigenstate expressions
are given, hence the results are free from the commonly adopted mass-insertion
approximation. In  the numerical analysis, we focus here only on the influence
of the trilinear $\l'_{ijk}$  couplings and report on the possibility of a few orders 
of magnitude improvement for the bounds on a few combinations of the $\l'$ 
couplings. Our study shows that the Wilson coefficients of the current-current 
operators due to $R$-parity violation dominate over the  direct contributions to the 
penguins. However, the inter-play of various contributions is complicated
due to the QCD corrections which we elaborate here.
\end{abstract}
\maketitle
\newpage
\section{Introduction}
With the successful testing of the gauge sector during the LEP era,
the standard model (SM) of particle physics has certainly established
itself as an essentially correct theory at and below the GeV scale.
However, it still leaves many basic issues to be investigated.
Even if one forgets about the  hierarchy problem,  and the need for neutrino masses, 
a cursory look at the error bars of about 20 \% on the various flavor parameters 
\cite{pdg2}, and their wide range of apparently arbitrary numerical values suggests 
that the flavor sector is still much of a misery. Within flavor physics, it is the phenomena 
of the flavor changing neutral currents (FCNCs) that could be the window of the
physics beyond SM. Forbidden at the tree level due to the unitarity
of $V_{\mbox{\tiny CKM}}$, FCNCs are loop induced where exotic virtualities
(from supersymmetry or otherwise) could pop up and hence completely change the 
predictions for the decay rate. This could prove to be complementary to the direct search 
of the exotic particles by hinting toward a particular class of models
or by ruling out large regions in the parameter space of beyond SM models,
and thus, guiding the accelerator searches.

Among the FCNCs, the process \Bsg\ is particularly attractive, both theoretically as 
well as experimentally. From the point of view of
theory, though the calculation is very complicated and was a major challenge 
due to the ambiguities following `scheme dependence' of the decay rate
right at the leading log (LL) \cite{bsg_classics,cella}, it was worth an effort.
Unlike many hadronic processes, it can largely be freed from the
scale and scheme dependence within the framework of renormalization
group improved perturbation theory and heavy mass expansion together with the
assumption of quark-hadron duality. All this is possible because
$m_b >> \Lambda_{\ssc QCD}$ and hence, the inclusive decay \Bsg\  is very well
approximated by the corresponding partonic level transition \bsg\ 
where the non-perturbative pathogens play only sub-leading role
being suppressed by at least two powers of $m_b$
(the corrections are less than 10\%). The next to leading log (NLL)
result for the branching fraction in the case of SM 
is given as \cite{gam-mis-01,bur-mis-02}:
\label{br-th}
\be Br\left[B \rightarrow X_s + \g \;(E_{\g} > 1.6 GeV) \right]_{\ssc \mathrm{SM}} 
= (3.57 \pm 0.30) \times 10^{-4}\;.
\ee
Here, $E_{\g}$ is the energy cut on the photon spectrum to get rid of the
background photons. The above result implies that this particular `rare' decay is 
actually not so rare. This is because, unlike other rare decays, where the rate is
$\propto G^2_{\ssc F}\a^2_{\ssc QED}$, here it is $\propto G^2_{\ssc F}
\a_{\ssc QED}$. Also due to the heavy top in the loop, GIM suppression is not
much effective. The high rate has already been quite well measured by
CLEO \cite{cleo_bsg}, BELLE \cite{belle_bsg} and ALEPH \cite{aleph_bsg}.
The results of different experiments are consistent with each other. A weighted
average of the available experimental measurement is problematic, because the model
dependence errors (and also the systematic errors) are correlated and differ within
the various measurements. An analysis taking into account the correlations, leads to 
the following world average \cite{world_average}
\be
\label{br-exp}
Br\left[B \rightarrow X_s + \g \;(E_{\g} > 1.6 GeV)\right]_{\ssc \mathrm{EXP}} 
= (3.34 \pm 0.38) \times 10^{-4}\;.
\ee
Within 1$\s$ it matches the SM predictions \cite{gam-mis-01,bur-mis-02}.
So it is clear that there is not much room for new physics contributions\footnote{See the note added in proof at the end.}. 
Either the new physics contributions are negligible, or there are large
cancellations among various contributions.
Thus, a study of new physics contributions  can help to
reduce the viable regions in parameter space of the relevant model. This channel is 
indeed very well studied in the most popular theory of 
beyond SM physics, namely the  minimal supersymmetric 
standard model (MSSM) \cite{mssm,baer,carena}. There is already a vast literature on the 
topic and we are not adding to that here. The imposed baryon number (B) and 
lepton number (L) conservation (in the form of $R$-parity) within MSSM, is totally 
{\it ad hoc}. It is an overkill of a problem of proton decay which 
can be easily rescued by alternatives like baryon parity.
It also forbids the neutrino masses that would otherwise arise naturally.
In fact, super-symmetrizing the SM with  its sacred gauge symmetries 
and renormalizability, one naturally lands into the generic supersymmetric
standard model or the supersymmetry (SUSY) without $R$-parity \cite{otto-gssm}. 
However, due to the presence of a large number of free parameters, 
any phenomenological study of $R$-parity violation (RPV) models is a daunting task. 
On the other hand, one may consider the origin and suppression of these parameters 
as due to a spontaneously broken anomalous Abelian family symmetry which is 
invoked to understand the fermion mass hierarchies \cite{fn}. In ref.\cite{u1} such 
an attempt is made to understand the pattern of $R$-parity  violation consistent with
phenomenology at the electroweak scale. However, in the absence of a compelling
model of the kind, it is imperative that all possible phenomenological  constraints on 
these,{\it a priori}, free parameters be studied. Such complexity necessitates 
a careful examination of the exact definition of the RPV couplings, and the choice 
of an optimal parametrization of the latter, or rather the full model Lagrangian,
that can simplify the structure of mass matrices to enable an analytical
diagonalization in a suitable approximation. The single vev parametrization (SVP), first
explicitly advocated in \cite{otto-svp} is such a parametrization. In this paper, we 
investigate the  RPV contributions from the generic model to the process \bsg\
within the framework of SVP \cite{bsg-icfp}. We give analytical
formulae that include all possible contributions without {\it a priori} assumption on
the form of R-parity violation. In the same spirit as an earlier study on
$\m \rightarrow e + \g$ \cite{otto-mueg}, we give mass eigenstate expressions,
hence free from the commonly adopted mass-insertion approximation. In the
numerical analysis, we focus in this paper only on the influence of the 
trilinear $\l'_{ijk}$ couplings and report on the possibility of a few orders
of magnitude improvement for the bounds on a few combination of 
$\l'_{ijk}$ parameters. Our formulation also has the advantage that the physical
meaning of the $\l'$-couplings would not change at all, whether the numerical values
of any of the other RPV parameters are vanishingly small or otherwise.

There have been some studies on the process within the general framework of 
$R$-parity violation. More systematic analysis are exemplified by 
refs.\cite{carlos,besmer}. Ref.\cite{carlos},  fails to consider the
additional 18 four-quark operators which, in fact, give the dominant contribution
in most of the cases. As we shall demonstrate, the interplay of various contributions
along with the QCD corrections, is very complicated and hence various contributions
could add up or cancel depending on a particular case. The more recent work of
ref.\cite{besmer} has considered a complete operator basis. However, we find their
formula for Wilson coefficient incomplete, and they do not report on the possibility
of a few orders of magnitude improvement on the bounds for certain combinations of
RPV couplings, as we present here. We must mention here that around the same time 
as ref.\cite{besmer} authors of ref.\cite{chun-bsg} published an  analysis of the 
CP-asymmetries in the radiative B-decays, which involved a similar calculation 
machinery with the extended operator basis. 
To the best of our knowledge all the available studies on the topic in the
literature, in fact are incomplete or incorrect to some extent. This is mostly
a problem of incomplete consideration for the RPV couplings of the bilinear
type and the mass mixings the latter produce in the sectors of fermions and scalars.
We present here a complete treatment of the full operator basis together with the
corresponding Wilson coefficients under the truly generic and consistent formulation
of SUSY without $R$-parity. In fact, the existence of interesting contributions from
combinations of bilinear and trilinear RPV couplings in the process has been noted in 
the earlier studies of the model under the SVP formulation \cite{otto-jhep,kong-keum}. 
We present the full analytical result here. However, due to other complications involved,
we will postpone any numerical study on the aspect to a later publication. Focusing our
attention here on the trilinear RPV parameters, we perform our analysis at the leading 
log order. Such an approximation entails uncertainties of about 25 \%. However, owing 
to sensitive dependence of the result on the large number of input parameters, 
we consider it worth-while to examine the relevant parameter space and obtain 
order of magnitude bounds before performing any precision analysis. 

The paper is organized as follows: In section II we briefly summarize 
the main features of SVP parametrization and also set our notation.
Section III deals with the effective Hamiltonian formulation, the extended
operator basis and the Wilson coefficients. Section IV discusses
the decay rate calculation in relation to the anomalous dimension matrix and the 
renormalization group running of the full set of 28 operators. We discuss 
our major numerical results in section V,  and conclude the paper after. Some details
involving the full set of 28 operators and the $28 \times 28$ anomalous dimension 
matrix is left to an  appendix. The numerical results discussed here are restricted
to those obtained from combinations trilinear couplings along. Contributions
from combinations of a trilinear and a bilinear parameter, the latter going into
the loop diagrams through RPV mass mixings, form another class of novel results
essentially not studied before. We present that latter part, from our parallel study, in 
an independent publication\cite{017}.
\section{ The Single VEV Parametrization}
The most general renormalizable superpotential for the generic supersymmetric
SM (without $R$-parity) can be written  as
\bea
W \!\!\!\! &=& \!\!\!\!\varepsilon_{ab}\Big[ \mu_{\alpha}  
\hat{H}_u^a \hat{L}_{\alpha}^b 
+ h_{ik}^u \hat{Q}_i^a   \hat{H}_{u}^b \hat{U}_k^{\scriptscriptstyle C}
+ \lambda_{\alpha jk}^{\!\prime}  \hat{L}_{\alpha}^a \hat{Q}_j^b
\hat{D}_k^{\scriptscriptstyle C} 
 +\;
\frac{1}{2}\, \lambda_{\alpha \beta k}  \hat{L}_{\alpha}^a  
 \hat{L}_{\beta}^b \hat{E}_k^{\scriptscriptstyle C} \Big] + 
\frac{1}{2}\, \lambda_{ijk}^{\!\prime\prime}  
\hat{U}_i^{\scriptscriptstyle C} \hat{D}_j^{\scriptscriptstyle C}  
\hat{D}_k^{\scriptscriptstyle C}\; ,
\eea\normalsize
where  $(a,b)$ are $SU(2)$ indices, $(i,j,k)$ are the usual family (flavor) 
indices, and $(\a, \b)$ are the extended flavor indices going from $0$ to $3$.
In the limit where $\lambda_{ijk}, \lambda^{\!\prime}_{ijk},  
\lambda^{\!\prime\prime}_{ijk}$ and $\mu_{i}$  all vanish, 
one recovers the expression for the $R$-parity preserving case, 
with $\hat{L}_{0}$ identified as $\hat{H}_d$. Without $R$-parity imposed,
the latter is not {\it a priori} distinguishable from the $\hat{L}_{i}$'s.
Note that $\lambda$ is antisymmetric in the first two indices, as
required by  the $SU(2)$  product rules, as shown explicitly here with 
$\varepsilon_{\scriptscriptstyle 12} =-\varepsilon_{\scriptscriptstyle 21}=1$.
Similarly, $\lambda^{\!\prime\prime}$ is antisymmetric in the last two 
indices, from $SU(3)_{\scriptscriptstyle C}$. 
$R$-parity is exactly an {\it ad hoc} symmetry put in to make $\hat{L}_{0}$,
stand out from the other $\hat{L}_i$'s as the candidate for  $\hat{H}_d$.
It is defined in terms of baryon number, lepton number, and spin as, 
explicitly, ${\mathcal R} = (-1)^{3B+L+2S}$. The consequence is that 
the accidental symmetries of baryon number and lepton number in the SM 
are preserved, at the expense of making particles and super-particles having 
a categorically different quantum number, $R$-parity. As mentioned above, $R$-parity
hence kills the dangerous proton decay but also forbids neutrino masses
within the model.

After the supersymmetrization of the SM, however, some of the superfields lose the exact
identities they have in relation to the physical particles. The latter have to
be mass eigenstates, which have to be worked out from the Lagrangian of the model.
Assuming electroweak symmetry breaking, we have now five (color-singlet) charged
fermions, for example. There are also 1+4 VEV's admitted, together with a SUSY
breaking gaugino mass. If one writes down naively the (tree-level) mass matrix,
the result is extremely complicated 
with all the $\mu_{\scriptscriptstyle \a}$ and $\lambda_{\a\b k}$ couplings
involved, from which the only definite experimental data are the three physical
lepton masses as the light eigenvalues, and the overall magnitude of
the electroweak symmetry breaking VEV's. The task of analyzing the model seems
to be formidable.

Doing phenomenological studies without specifying a choice 
of flavor bases is, however, ambiguous. It is like doing SM quark physics with 18
complex Yukawa couplings, instead of the 10 real physical parameters.
As far as the SM itself is concerned, the extra 26 real parameters
are simply redundant, and attempts to relate the full 36 parameters to experimental 
data will be futile. In the case at hand, the choice of an optimal
parametrization mainly concerns the 4 $\hat{L}_\alpha$ flavors. In the
single vev parametrization\cite{otto-svp} (SVP), flavor bases 
are chosen such that : 
1) among the $\hat{L}_\alpha$'s, only  $\hat{L}_0$, bears a VEV,
{\it i.e.} {\small $\langle \hat{L}_i \rangle \equiv 0$};
2)  {\small $y^{e}_{jk} (\equiv \lambda_{0jk}) 
=\frac{\sqrt{2}}{v_{\scriptscriptstyle 0}} \,{\rm diag}
\{m_{\scriptscriptstyle 1},
m_{\scriptscriptstyle 2},m_{\scriptscriptstyle 3}\}$};
3) {\small $y^{d}_{jk} (\equiv \lambda^{\!\prime}_{0jk} =-\lambda_{j0k}) 
= \frac{\sqrt{2}}{v_{\scriptscriptstyle 0}}\,{\rm diag}\{m_d,m_s,m_b\}$}; 
4) {\small $h^{u}_{ik}=\frac{\sqrt{2}}{v_{\scriptscriptstyle u}}
V_{\mbox{\tiny CKM}}^{\!\scriptscriptstyle T} \,{\rm diag}\{m_u,m_c,m_t\}$}, where 
${v_{\scriptscriptstyle 0}} \equiv  \sqrt{2}\,\langle \hat{L}_0 \rangle$
and ${v_{\scriptscriptstyle u} } \equiv \sqrt{2}\,
\langle \hat{H}_{u} \rangle$. Thus, the parametrization singles out the
$\hat{L}_0$ superfield as the one containing the Higgs. As a result,
it gives the complete RPV effects on the {tree-level mass matrices} 
of all the states (scalars and fermions) the simplest structure.
The latter is a strong technical advantage.

\par The soft SUSY breaking part of the Lagrangian can be written as 
\bea
 \!\!\!\!\!\!\!\! V_{\rm soft}
&=& \tilde{Q}^\dagger \tilde{m}_{\!\scriptscriptstyle {Q}}^2 \,\tilde{Q} 
+\tilde{U}^{\dagger} 
\tilde{m}_{\!\scriptscriptstyle {U}}^2 \, \tilde{U} 
+\tilde{D}^{\dagger} \tilde{m}_{\!\scriptscriptstyle {D}}^2 
\, \tilde{D} +
 \tilde{L}^\dagger \tilde{m}_{\!\scriptscriptstyle {L}}^2  \tilde{L}  
+
 \tilde{E}^{\dagger} \tilde{m}_{\!\scriptscriptstyle {E}}^2 
\, \tilde{E}
+ \tilde{m}_{\!\scriptscriptstyle H_{\!\scriptscriptstyle u}}^2 \,
|H_{u}|^2 
+  \Big[ \,
\frac{M_{\!\scriptscriptstyle 1}}{2} \tilde{B}\tilde{B}
   + \frac{M_{\!\scriptscriptstyle 2}}{2} \tilde{W}\tilde{W}
 \nonumber \\
&&+
\frac{M_{\!\scriptscriptstyle 3}}{2} \tilde{g}\tilde{g}
+\epsilon_{\!\scriptscriptstyle ab} \Big( \,
  B_{\a} \,  H_{u}^a \tilde{L}_\a^b 
+ A^{\!\scriptscriptstyle U}_{ij} \, 
\tilde{Q}^a_i H_{u}^b \tilde{U}^{\scriptscriptstyle C}_j 
+
 A^{\!\scriptscriptstyle D}_{ij} 
H_{d}^a \tilde{Q}^b_i \tilde{D}^{\scriptscriptstyle C}_j  
+ A^{\!\scriptscriptstyle E}_{ij} 
H_{d}^a \tilde{L}^b_i \tilde{E}^{\scriptscriptstyle C}_j 
+ A^{\!\scriptscriptstyle \lambda^\prime}_{ijk} 
\tilde{L}_i^a \tilde{Q}^b_j \tilde{D}^{\scriptscriptstyle C}_k
\nonumber \\
&&+ 
\frac{1}{2}\, A^{\!\scriptscriptstyle \lambda}_{ijk} 
\tilde{L}_i^a \tilde{L}^b_j \tilde{E}^{\scriptscriptstyle C}_k  \Big)
+ \frac{1}{2}\, A^{\!\scriptscriptstyle \lambda^{\prime\prime}}_{ijk}
 \tilde{U}^{\scriptscriptstyle C}_i  \tilde{D}^{\scriptscriptstyle C}_j  
\tilde{D}^{\scriptscriptstyle C}_k  
  +  \mbox{\normalsize h.c.} \Big]\;,
\label{soft}
\eea
where we have separated the $R$-parity conserving $A$-terms from the 
RPV ones (recall $\hat{H}_{d} \equiv \hat{L}_0$). Note that 
$\tilde{L}^\dagger \tilde{m}_{\!\scriptscriptstyle \tilde{L}}^2  \tilde{L}$,
unlike the other soft mass terms, is given by a 
$4\times 4$ matrix. Explicitly, 
$\tilde{m}_{\!\scriptscriptstyle {L}_{00}}^2$ corresponds to 
$\tilde{m}_{\!\scriptscriptstyle H_{\!\scriptscriptstyle d}}^2$ 
of the MSSM case while 
$\tilde{m}_{\!\scriptscriptstyle {L}_{0k}}^2$'s give RPV mass mixings.
Going from here, it is straight forward to obtain the squark and slepton masses. 
No RPV A-term enters the mass matrices for instance. Full matrices have been
worked out, together with various diagonalizing matrix elements from the 
perturbative approximations for the matrices within the experimentally viable
setting of small RPV couplings. Readers are referred to ref.\cite{otto-gssm}
for further details.
\section{The Effective Hamiltonian}
\subsection{The basic Strategy}
The partonic transition \bsg\ is described by the magnetic penguin diagram
(see Fig.\ref{higgs1}). QCD corrections are obtained by attaching a gluon 
line to the quarks lines. Such a correction forms a power series in
$\a_{\mbox{\tiny QCD}}\, (M_{\!\ssc W})\,\mathrm{ln}(M^{2}_{\!\ssc W}/m^2_b) \sim 0.7$ 
which is too large to be treated in the perturbation framework. This can be efficiently 
handled within the framework of operator product expansion (OPE) and the 
renormalization group (RG) improved perturbation theory \cite{buras-review}.
One writes down the effective Hamiltonian (${\cal H}_{{\rm eff}}$) as a series of local 
operators ${\cal Q}_{{\sl i}}(\m)$ multiplied by the corresponding Wilson 
coefficients $C_i (\m)$. Both, the operators and the Wilson coefficients are function 
of the scale $\m$, however, the effective Hamiltonian is expected to be scale 
independent. Thus, we can write
\be
{\cal H}_{{\rm eff}}\, =\,  \sum_i\,  C_i(\m)\, {\cal Q}_{{\sl i}}(\m)\;.
\ee
The choice of scale $\m$ is arbitrary, and separates high scale physics ($>\m$) 
and the low scale ($<\m$). It is typically taken to be the mass scale
of the decaying particle (in this case the $b$-quark). There are three major steps in 
the evaluation of the decay rate. They are:
\begin{figure}
\includegraphics[scale=0.6]{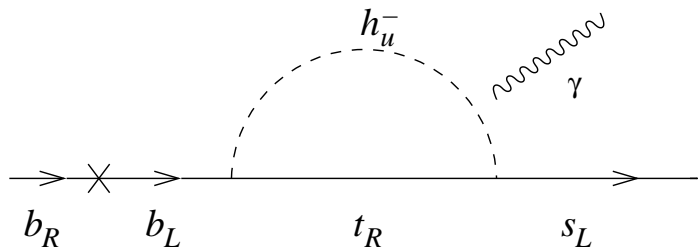}
\caption{\label{higgs1} 
Magnetic penguin diagram with Higgs in the loop. Here the chirality flip is
on the external $b$-quark. Note that the photon can be attached to any 
electrically charged particle inside the loop but not outside the loop.
Diagram with a photon outside the loop does not contribute to the
magnetic transition. There are similar diagrams with other SUSY particles
in the loop. For the chromo-magnetic penguin one attaches a gluon, instead of a 
photon to the quarks in the loop.}
\end{figure}
\begin{figure}
\includegraphics[scale=0.8]{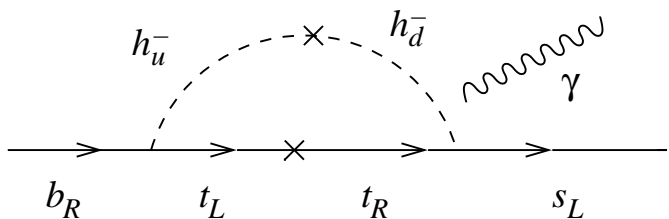}
\caption{\label{higgs2} 
A magnetic penguin digram but with the chirality flip
inside the loop.}
\end{figure}
\begin{itemize}
\item {\bf Matching at $M _{\!\ssc W}$:} $C_i$ are first evaluated perturbatively 
at the scale at which one decouples the heavy modes (say at $\m = M _{\!\ssc W}$).
This is done by imposing the equality of the effective, and the underlying theory Green
functions, at external momenta that are much smaller than the masses of the 
decoupled particles.
\item {\bf Resummation:} Since the theory contains two disparate scales $m_b$ 
and $M _{\!\ssc W}$, the large logarithms need to be resumed using RG improved 
perturbation theory. The RG equation for the $C_i$ are given as:
\be
\label{rge}
\frac{d C_i}{dln\m} = \g_{ji}C_j \;.
\ee
Here $\g_{ji}$ is the anomalous dimension matrix (ADM). It is obtained from the 
ultra-violet (UV) divergences in the theory. The derivation of ADM is the most difficult 
part of the calculation. Solving the RG equation one gets the $C_i$ at the 
scale $m_b$. Note that in our case the resummation is performed at the LL order.
\item {\bf Matrix elements:} Having obtained the $C_i(m_b)$ one finds the hadronic
matrix elements of the relevant operators $\lag\cq_i\rag$. Typically this involves some
non-perturbative method; but in case of inclusive B decays, non-perturbative corrections
are sub-leading and one can approximate the decay by a corresponding partonic level
transition.
\end{itemize}
\subsection{The Operators}
Having outlined the strategy, let us list the army of operators beginning with the 
SM \cite{bsg_classics}. So far as one neglects the QCD corrections, the relevant operator is the 
magnetic penguin. However, once the penguin is dressed with the gluonic lines, it 
mixes with other current-current and QCD-penguin operator. They are given as:
\bea
\label{operators}
{\cal Q}_{1,2}&\, =\,& \left({\bar s_{L\a}}\,\g^{\m}\,
b_{L\a,\b}\right)\,\left({\bar c_{L\b}}\,\g_{\m}\, 
c_{L\b ,\a}\right)\;,  \\
{\cal Q}_{3,4}&\,=\,& \left({\bar s_{L\a}}\,\g^{\m}\,b_{L\a,\b}\right)\,
\sum_{i=u,c,d,s,b}\,\left(\bar{q}_{Li\b}\,\g_{\m}\,q_{Li\b \a}\right)\;, \\
{\cal Q}_{5,6}&\,=\,& \left({\bar s_{L\a}}\,\g^{\m}\,b_{L\a,\b}\right)\,
\sum_{i=u,c,d,s,b}\,\left(\bar{q}_{Ri\b}\,\g_{\m}\,q_{Ri\b ,\a}\right)\;, \\
{\cal Q}_{\sl 7}&\, =\,& \frac{e}{16 \pi^2}\,m_b\, {\bar s_{L\a}}
\,\s_{\m\n}\,b_{R\a}\, F^{\m\n}\;, \\
{\cal Q}_{8}&\,=\,&\frac{g_s}{16 \pi^2}\,m_b\, {\bar s_{L\a}}
\,\s_{\m\n}\,b_{R\b}\,t^{a\a\b} G^{a\m\n}\;.
\eea
Here, we have used Fierz re-ordering to write the first two operators. $\a$ and $\b$ 
are the colour indices and $t^a$ are the generators of $SU(3)$. Other notations are 
self explanatory. Operators ${\cal Q}_3$ to ${\cal Q}_6$ are the
QCD penguins, and $\cq_{7,8}$ are the magnetic and the chromomagnetic penguins
respectively involving up-type quark and W-boson. We would like to remark here that 
for a novice there is some chance of confusion regarding the choice of operator basis. 
The above mentioned operator basis is arrived at by considering the external states 
on-shell and hence these cannot be transformed into each other using equations of 
motion \cite{politzer}. In an off-shell calculation one must consider additional operators 
\footnote{For instance see the ref.\cite{cella} for an off-shell basis. Ref.\cite{simma}
gives a careful treatment about the issues regarding equations of motion and
effective Lagrangian, often neglected in other works.}. 
Even for the on-shell basis there exist one more basis choice in the literature introduced 
by the authors of \cite{cmm}
\footnote{The operators are written in such a manner
that one does not encounter ill-defined trace of the product of 
odd number of gamma matrices including a $\g_5$. This allows them to use the fully
anti-commuting $\g_5$ in dimensional regularization.}.
\par Since the magnetic penguins are chirality 
flipping operators, they are necessarily proportional to the fermionic mass term. 
In principle one can also write down the following two operators obtained by
flipping the chirality of $\cq_{7,8}$. 
\bea
{\widetilde{\cal Q}}_{\sl 7}&\, =\,& \frac{e}{16 \pi^2}\,m_s\, {\bar s_{R\a}}
\,\s_{\m\n}\,b_{L\a}\, F^{\m\n}\;, \\
{\widetilde{\cal Q}}_{8}&\,=\,&\frac{g_s}{16 \pi^2}\,m_s\, {\bar s_{R\a}}
\,\s_{\m\n}\,b_{L\b}\,t^{a\a\b} G^{a\m\n}\;.
\eea
However, these are suppressed by a factor of $m_s/m_b$ and hence generally 
neglected in the SM analysis. They do give significant contribution in 
case of SUSY without $R$-parity under the discussion in this paper though.
MSSM contributions are very well studied in the literature. We shall make only
brief remarks over here and refer the reader to the literature \cite{mssm,baer,carena} 
for a detailed
discussion. In MSSM, over and above the usual, W-t loop contributions, one can have 
charged Higgs-up-squarks, chargino-up-squarks, gluino-down squarks and neutralino 
down-squarks contributions. Since there is left-right scalar mixing in the squark sector, 
there is a possibility of a mass-insertion inside the loop (see for instance 
Fig.\ref{higgs2} as example with a mass-insertion inside the loop).
Since these are mass-terms of the already decoupled heavy modes, they are included 
in the corresponding Wilson coefficients that encode the dynamics of heavy modes. So
in the MSSM, there are new contributions to the Wilson coefficient of $\cq_7$, the 
operator basis is usually assumed to be the same as in SM. However, this is not quite 
true. It has been demonstrated in \cite{borz-gluino} that there could be 100 extra 
scalar, vector and tensor-type four quark operators, formally higher order in strong 
coupling (compared to the SM operators). The 40 vector type operators are 
generated by two gluino box diagrams and gluino-gluon penguin diagrams. However, 
the vector four-quark operators do not mix with the SM operators at one-loop and hence 
contribute only at the next to leading log. The 60 scalar and tensor type operators are 
generated by the two gluino box diagram. These could in-principle mix with the 
$\cq_{7,8}$. However, this mixing turns out to be sub-dominant.
The scalar and tensor type four-quark operators mix only with the dimension-6 penguins
(corresponding to the chirality flip on the external quark) and hence their contribution to
the amplitude is suppressed by $m_b/m_{\tilde g}$ relative to that of dimension-5 ones 
\cite{okamura}. So we shall neglect all this extra 100 operators.

Giving up the restriction of $R$-parity adds to the list of four quark operators. 
From an inspection of the super-potential it is straightforward to write down the relevant 
operators. It is clear that there cannot be any contribution from the $\l$ couplings.
One gets 12 new four quark operators from the combinations of two $\l'$-couplings and 
6 new four-quark operators from the combinations of two $\l''$-couplings. 
With so many extra operators of different origin, a notation that is intuitively suggestive, 
compact, unambiguous and at the same time offers convenient expression of summation 
in formulas is virtually impossible. We make a simple choice for the notation as follows:
Keeping the notation for SM operators as a yard-stick, we divide
the operators according to their chiralities (the advantage of such a 
classification would soon become clear). Those operators that lead to
$b_{\!\ssc R}\rightarrow s_{\!\ssc L}$ transition are appended to the list of SM 
operators whereas the operators leading to $b_{\!\ssc L} \rightarrow s_{\!\ssc R}$ 
transition are denoted with a tilde ({\it i.e.,} in the form ${\wtl {\cal Q}_i}$) and form 
a new list. Obviously the above two classes of operators are related by 
$L\rightarrow R$ flip. This choice is not without potential ambiguity. For instance, the 
notation $\cq_9,\cq_{10}$ typically denotes semi-leptonic operators, but since we are 
not dealing with semi-leptonic process in this paper, there should not be any ambiguity.
With these comments we write down the additional operators obtained from RPV. 

\vspace{0.3cm}
\noindent{\bf $\l'\l'$ case:}
\bea
{\cal Q}_{9-11}& =& \left({\bar s}_{L\a}\,\g^{\m} \, b_{L\b}\right)
\, \left({\bar q}_{R\b}\,\g_{\m}\,q_{R\a}\right)~;~q = d,s,b \;,  \\
{\widetilde{\cal Q}}_{ 9-13} & = & \left({\bar s}_{R\a}\,\g^{\m}\,b_{R\b}\right)
\, \left({\bar q}_{L\b}\, \g_{\m}\,q_{L\a}\right)~;~ q\, = d,s,b,u,c\;,  \\
{\widetilde {\cal Q}_{3,4}} &=& \left({\bar s}_{R\a}\,\g^{\m}b_{R\a,\b}\right)
\,\sum_{i=u,c,d,s,b}\,\left({\bar q}_{Ri\b}\,\g^{\m}\,q_{Ri\b,\a}\right)\;,  \\
{\widetilde {\cal Q}_{5,6}} &=& \left({\bar s}_{R\a}\,\g^{\m}b_{R\a,\b}\right)
\,\sum_{i=u,c,d,s,b}\,\left({\bar q}_{Li\b}\,\g^{\m}\,q_{Li\b,\a}\right)\;.
\eea

\vspace{0.3cm}
\noindent{\bf $\l''\l''$ case:}
\bea
{\widetilde {\cal Q}}_{1,2} & =&  \left({\bar s}_{R\a}\,\g^{\m}\,b_{R\a,\b}\right)
\,\left({\bar c}_{R\b}\,\g_{\m}\,c_{R\b,\a}\right)\;, \\ 
{\widetilde {\cal Q}}_{14,15} & =&  \left({\bar s}_{R\a}\,\g^{\m}\,b_{R\a,\b}\right)
\,\left({\bar u}_{R\b}\,\g_{\m}\,u_{R\b,\a}\right)\;, \\ 
{\widetilde {\cal Q}}_{16,17} & =&  \left({\bar s}_{R\a}\,\g^{\m}\,b_{R\a,\b}\right)
\,\left({\bar d}_{R\b}\,\g_{\m}\,d_{R\b,\a}\right)\;. 
\eea
Thus the total number of operators becomes 28 which mix upon QCD renormalization.

\subsection{Wilson Coefficients}
The most important operator for the decay \bsg\ is of course the magnetic penguin
$\cq_7$. In order to obtain the corresponding $C_7$ we first need to define
the relevant interaction Lagrangian. At this point we would like to emphasize
that since we are working in the most general setting of RPV, there is an important
distinction in the way we write down the interaction Lagrangian and the Wilson 
coefficients as compared to ref.\cite{besmer}. Our analytical formulas contain all 
possible contributions from RPV including the bilinear RPV couplings.
In ref.\cite{besmer} the treatment of bilinear (RPV) parameters is slightly erroneous. 
They work in a basis where bilinears ($\m_i$'s) have been rotated away. 
This {\it does not} get rid of all the RPV effects beyond those described by the
trilinear RPV couplings. RPV mass mixings are still to be found among the fermions
and the scalars \cite{otto-gssm}. In fact such mass mixing effects through the
``sneutrino'' VEVs, render the definition of the flavor basis of the charged leptons
and the down-sector quarks in the formulation of the Lagrangian ambiguous 
\cite{otto-gssm}. The problem is totally avoided in our formulation. Our formulae are
given in terms of exact mass eigenstates, instead of the often used mass insertion
approximation. Because of the RPV mass mixings our formulae for the Wilson
coefficients do not factor into the MSSM contributions and RPV contributions as
treated in ref.\cite{besmer}. With these remarks we list here the interaction Lagrangian.
\vskip 8pt
\noindent{\bf Gluino-quark-squark vertex}\\
\be
\label{gluino-vert}
{\cal L}^{\tilde{g}} = {g_{s}} 
\overline{\Psi}(\tilde{g}) \Phi^{\dagger} (\tilde d_m)
\left[ {\cal G}^{\scriptscriptstyle R}_{\ssc mi} 
{1 + \gamma_{\scriptscriptstyle 5} \over 2}  + 
{\cal G}^{\scriptscriptstyle L}_{\ssc mi} 
{1 - \gamma_{\scriptscriptstyle 5} \over 2}  \right] 
{\Psi}({d_{i}}) 
\, +\, \mbox{h.c.}\;,
\ee
where,
\be
{\cal G}^{{\scriptscriptstyle L}^*}_{\scriptscriptstyle mi}
= -\sqrt{8 \over 3} \, {\cal D}^{d}_{\!im}\;,
\qquad\qquad
{\cal G}^{{\scriptscriptstyle R}^*}_{\scriptscriptstyle mi} 
= \sqrt{8\over 3} \, {\cal D}^{d}_{\!(i+3)m} \; ,
\ee
with ${\cal D}^{d^{\dagger}}{\cal M}^2_{\mbox{\tiny{D}}}{\cal D}^d
= diag\{{\cal M}^2_{\mbox{\tiny{D}}}\}$ while ${\cal M}^2_{\mbox{\tiny{D}}}$
is $6\times 6$ being the down-squark mass-matrix.
Here $i,j,k=1$ to $3$ (same below), $m=1$ to $6$ for the scalar (squark) mass
eigenstates.
\vskip 8pt
\noindent{\bf Chargino(charged-lepton)-quark-squark vertices}\\
\be
\label{chargino-vert}
{\cal L}^{\!\chi^{\mbox{-}}} = {g_{\scriptscriptstyle 2}} 
\overline{\Psi}(\chi_n^{\mbox{-}}) \Phi^{\dagger}({\tilde{u}_{m}})
\left[ {\cal C}^{\scriptscriptstyle L}_{\ssc nmi} 
{1 - \gamma_{\scriptscriptstyle 5} \over 2}  + 
{\cal C}^{\scriptscriptstyle R}_{\ssc nmi} 
{1 + \gamma_{\scriptscriptstyle 5} \over 2}  \right]  
{\Psi}({d_{i}}) 
\ + \mbox{h.c.}\;,
\ee
where,
\bea
{\cal C}^{{\!\scriptscriptstyle L}^*}_{\scriptscriptstyle nmi}
&=& - \mbox{\boldmath $V$}_{\!\!1n} \,{\cal D}^{u}_{\!im} 
+ {y_{\!\scriptscriptstyle u_p} \over g_{\scriptscriptstyle 2} } \,
V_{\!\mbox{\tiny CKM}}^{pi^*} \, 
\mbox{\boldmath $V$}_{\!\!2n} \, {\cal D}^{u}_{\!(p+3)m}\;,
\nonumber \\
{\cal C}^{{\!\scriptscriptstyle R}^*}_{\scriptscriptstyle nmi} 
&=&  {y_{\!\scriptscriptstyle d_i} \over g_{\scriptscriptstyle 2} }\,
 V_{\!\mbox{\tiny CKM}}^{hi^*} \,
\mbox{\boldmath $U$}_{\!2n} \, {\cal D}^{u}_{\!hm}
+ {\lambda^{\!\prime}_{qri} \over g_{\scriptscriptstyle 2} }\,
 V_{\!\mbox{\tiny CKM}}^{hr^*} \,
\mbox{\boldmath $U$}_{\!(q+2)n} \, {\cal D}^{u}_{\!hm}\;. 
\eea
Here, and throughout the paper, the $m$ index counts the scalar mass eigenstates
and the $n$ index that of the fermions.
Matrix ${\cal D}^u$ diagonalizes 
the $6\times6$ up-squark mass matrix {\it i.e.} 
${\cal D}^{u^{\ssc \dagger}}{\cal M}^2_U{\cal D}^u = \rm{diag}\{{\cal M}^2_U\}$,
and $m = 1$ to $6$ for the up-squarks. 
Similarly 
$\mbox{\boldmath $V$}^{\dagger}{\cal M}_C \mbox{\boldmath $U$}
=\rm{diag}\{{\cal M_C}\}$ for the $n=1$ to $5$ charged fermions
(charginos and charged leptons).
\vskip 8pt
\noindent{\bf Neutralino(neutrino)-quark-squark vertices}\\
\be
\label{neutralino-vert}
{\cal L}^{\!\chi^{\mbox{\tiny 0}}} = {g_{\scriptscriptstyle 2}} 
\overline{\Psi}(\chi_n^{\mbox{\tiny 0}}) \Phi^{\dagger}({\tilde{d}_{m}})
\left[ {\cal N}^{\scriptscriptstyle L}_{\ssc nmi} 
{1 - \gamma_{\scriptscriptstyle 5} \over 2}  + 
{\cal N}^{\scriptscriptstyle R}_{\ssc nmi} 
{1 + \gamma_{\scriptscriptstyle 5} \over 2}  \right]  
{\Psi}({d_{i}}) 
\ + \mbox{h.c.}\;,
\ee
where,
\bea
{\cal N}^{{\!\scriptscriptstyle L}^*}_{\scriptscriptstyle nmi}
\!\!\! &=& \! -\sqrt{2} \left\{ \tan\!\theta_{\!\scriptscriptstyle W} 
({\cal Q}_{\!d} -T_{3f}) \mbox{\boldmath $X$}_{\!\!1n}^*
+ T_{3f} \mbox{\boldmath $X$}_{\!\!2n}^* \right\} {\cal D}^{d}_{\!im} - 
{y_{\!\scriptscriptstyle d_i} \over g_{\scriptscriptstyle 2} } 
\mbox{\boldmath $X$}_{\!\!4n}^* \, {\cal D}^{d}_{\!(i+3)m} \nonumber\\ 
~&~& - {\lambda_{hip}^{\!\prime*} \over g_{\scriptscriptstyle 2} } 
\mbox{\boldmath $X$}_{\!\!(h+4)n}^*  {\cal D}^{d}_{\!(p+3)m} \;,
 \\
{\cal N}^{{\!\scriptscriptstyle R}*}_{\scriptscriptstyle nmi} 
\!\!\! &=& \! \sqrt{2} \, \tan\!\theta_{\!\scriptscriptstyle W} \,
{\cal Q}_{\!d} \, \mbox{\boldmath $X$}_{\!\!1n} \, {\cal D}^{d}_{\!(i+3)m}
- {y_{\!\scriptscriptstyle d_i} \over g_{\scriptscriptstyle 2} } \, 
\mbox{\boldmath $X$}_{\!\!4n} \, {\cal D}^{d}_{\!im}
- {\lambda_{hqi}^{\!\prime} \over g_{\scriptscriptstyle 2} } \,
\mbox{\boldmath $X$}_{\!\!(h+4)n} \, {\cal D}^{d}_{\!qm}\; ,
\eea
where $T_{3f}=-{1 \over 2}$. 
The neutral fermions indexed by $n$ have the $7\times7$ (neutralino-neutrino)
mass matrix to be diagonalized by X.
\vskip 8pt 
\noindent{\bf Charged Higgs(slepton)-quark-quark vertices}\\
\be
\label{selectron-vert}
{\cal L}^{\!\phi^{\mbox{-}}} = {g_{\scriptscriptstyle 2}} 
\overline{\Psi}(u_n)\Phi^{\dagger}(\phi_m^{{\mbox{-}}}) 
\left[ 
\widetilde{\cal C}^{\!\scriptscriptstyle L}_{\scriptscriptstyle nmi}
{1 - \gamma_{\scriptscriptstyle 5} \over 2}  + 
\widetilde{\cal C}^{\!\scriptscriptstyle R}_{\scriptscriptstyle nmi}
{1 + \gamma_{\scriptscriptstyle 5} \over 2}  \right]  
{\Psi}({d_{i}}) 
\ + \mbox{h.c.}\;,
\ee
where,
\bea
\widetilde{\cal C}^{{\!\scriptscriptstyle L}^*}_{\scriptscriptstyle nmi}
&=&  \frac{y_{\!\scriptscriptstyle u_n}}{g_{\scriptscriptstyle 2}} \,
  V_{\!\mbox{\tiny CKM}}^{ni^*} \, {\cal D}^{l}_{1m}  \;,
\nonumber \\
\widetilde{\cal C}^{{\!\scriptscriptstyle R}^*}_{\scriptscriptstyle nmi} 
&=&  \frac{y_{\!\scriptscriptstyle d_i}}{g_{\scriptscriptstyle 2}} \,
   V_{\!\mbox{\tiny CKM}}^{ni^*} \, {\cal D}^{l}_{2m} 
 +  \frac{\lambda_{jki}^{\!\prime}}{g_{\scriptscriptstyle 2}} \, 
  V_{\!\mbox{\tiny CKM}}^{nk^*} \, {\cal D}^{l}_{(j+2)m}  \; .
\eea
 ${\cal D}^l$ being the diagonalizing matrix for the $8\times 8$ charged-slepton 
charged-Higgs mass matrix.
\vskip 8pt
\noindent{\bf Neutral scalar(sneutrino)-quark-quark vertices}\\
\be
\label{sneutrino-vert}
{\cal L}^{\!\phi^{\mbox{\tiny 0}}} = {g_{\scriptscriptstyle 2}} 
\overline{\Psi}(d_n)\Phi^{\dagger}(\phi_m^{\mbox{\tiny 0}})
\left[ 
\widetilde{\cal N}^{\!\scriptscriptstyle L}_{\scriptscriptstyle nmi}
{1 - \gamma_{\scriptscriptstyle 5} \over 2}  + 
\widetilde{\cal N}^{\!\scriptscriptstyle R}_{\scriptscriptstyle nmi}
{1 + \gamma_{\scriptscriptstyle 5} \over 2}  \right]  
{\Psi}({d_{i}}) 
\ + \mbox{h.c.}\;,
\ee
where
\bea
 \widetilde{\cal N}^{{\!\scriptscriptstyle L}^*}_{\scriptscriptstyle nmi}
& =&  - \frac{y_{\!\scriptscriptstyle d_i}}{g_{\scriptscriptstyle 2}} \,
\delta_{in} \,
{1 \over \sqrt{2}} \,
[ {\cal D}^s_{\!2m} - i \, {\cal D}^s_{\!7m} ]
- \frac{\lambda_{pin}^{\!\prime*}}{g_{\scriptscriptstyle 2}} \,
{1 \over \sqrt{2}} \,
[ {\cal D}^s_{\!(p+2)m} -  i \, {\cal D}^s_{\!(p+7)m} ] \; ,
\nonumber \\
 \widetilde{\cal N}^{{\!\scriptscriptstyle R}^*}_{\scriptscriptstyle nmi}
& = &  
- \frac{y_{\!\scriptscriptstyle d_i}}{g_{\scriptscriptstyle 2}} \,
\delta_{in} \,
{1 \over \sqrt{2}} \,
[ {\cal D}^{s}_{\!2m} + i \, {\cal D}^{s}_{\!7m} ]
- {\lambda_{qni}^{\!\prime} \over g_{\scriptscriptstyle 2} } \, 
{1 \over \sqrt{2}} \,
[ {\cal D}^{s}_{\!(q+2)m} + i \, {\cal D}^{s}_{\!(q+7)m} ] \; ,
\eea
 ${\cal D}^s$ being the diagonalizing matrix for the $10\times 10$ neutral 
scalar mass matrix.

Next we present all the Wilson coefficients at the scale $M_{\!\ssc W}$. The most 
important is of course $C_7$. It can be decomposed into various contributions as follows.
\be
C_7 = C^{\ssc W}_7 \,+ \,C^{\ssc \tilde g}_7 \, + \,C^{\ssc \chi^-}_7 \, 
+\, C^{\ssc \chi^{\mbox{\tiny 0}}}_7 \, +\, C^{\ssc \phi^-}_7 \,
+\, C^{\phi^{\mbox{\tiny 0}}}_7\;.
\ee
The terms from left to right are the SM, gluino, chargino(charged-lepton), 
neutralino(neutrino), charged scalar(sleptons and Higgs) and neutral scalar(sneutrino) 
contributions. All except the last one are common to MSSM too. Note, however, that in
accordance with the above formulation, we are putting together in
$C^{\ssc \chi^-}_7$ and $C^{\ssc \chi^{\mbox{\tiny 0}}}_7$ contributions
from the light fermion mass eigenstates, namely the charged leptons and
neutrinos. Below we list the individual contributions for a decay 
$d_j \rightarrow d_i + \g$. 
($j =3$ and $i = 2$ for \bsg\ ) .
\bea
C^{\ssc W}_7 &=&
\frac{3g_2^2}{8}
{m_{\!\scriptscriptstyle u_n}^2 \over M_{\!\scriptscriptstyle W}^4} 
V_{\!\mbox{\tiny CKM}}^{nj}  V_{\!\mbox{\tiny CKM}}^{ni^*} 
 \left[
-F_2\!\!\left({m_{\!\scriptscriptstyle u_n}^2 \over 
M_{\!\scriptscriptstyle W}^2} \right) 
\!-\! {\cal Q}_{\!u} 
F_1\!\!\left({m_{\!\scriptscriptstyle u_n}^2 \over M_{\!\scriptscriptstyle W}^2}\right)  
\right]\;, \\
C^{\ssc \tilde g}_7 &=&
{2g^2_s \over 3} \;
{{\cal Q}_{\!\tilde{d}} \over M_{\!\scriptscriptstyle \tilde{d}_m}^2} 
\left[{\cal G}^{\scriptscriptstyle R}_{\scriptscriptstyle mj} 
{\cal G}^{\scriptscriptstyle L^*}_{\scriptscriptstyle mi} \;
{{M}_{\tilde{g}} \over m_{\!\scriptscriptstyle {d}_j}} \;
F_4\!\!\left({{M}_{\tilde{g}}^2 \over M_{\!\scriptscriptstyle \tilde{d}_m}^2} \right) \; 
 +
{\cal G}^{\scriptscriptstyle L}_{\scriptscriptstyle mj} 
{\cal G}^{\scriptscriptstyle L^*}_{\scriptscriptstyle mi} 
F_2\!\!\left({{M}_{\tilde{g}}^2 \over M_{\!\scriptscriptstyle \tilde{d}_m}^2} \right)
\right]\;,\\
C^{\ssc \chi^-}_7 &=&
\frac{{\cal C}^{\!\scriptscriptstyle R}_{\!\scriptscriptstyle nmj} 
{\cal C}^{\!\scriptscriptstyle L^*}_{\!\scriptscriptstyle nmi}}
{4M_{\!\scriptscriptstyle \tilde{u}_m}^2}\left[ 
{{M}_{\!\scriptscriptstyle \chi^{\mbox{-}}_n} \over 
{m}_{\!\scriptscriptstyle d_j}}
 {\cal Q}_{\!\tilde{u}}  
F_4\!\!\left({{M}_{\!\scriptscriptstyle \chi^{\mbox{-}}_{n}}^2 \over 
M_{\!\scriptscriptstyle \tilde{u}_m}^2} \right) 
+F_3\!\!\left({{M}_{\!\scriptscriptstyle \chi^{\mbox{-}}_{n}}^2 \over 
M_{\!\scriptscriptstyle \tilde{u}_m}^2} \right) 
\right] 
\!+\!
\frac{{\cal C}^{\!\scriptscriptstyle L}_{\!\scriptscriptstyle nmj} 
{\cal C}^{\!\scriptscriptstyle L^*}_{\!\scriptscriptstyle nmi}}
{4M_{\!\scriptscriptstyle \tilde{u}_m}^2}\!\!
\left[ {\cal Q}_{\!\tilde{u}} \; 
F_2\!\!\left({{M}_{\!\scriptscriptstyle \chi^{\mbox{-}}_{n}}^2 \over 
m_{\!\scriptscriptstyle \tilde{u}_m}^2} \right) 
+ F_1\!\!\left({{M}_{\!\scriptscriptstyle \chi^{\mbox{-}}_{n}}^2 \over 
M_{\!\scriptscriptstyle \tilde{u}_m}^2} \right) 
\right]\;,  \\
C^{\chi^{\mbox {\tiny 0}}}_7 &=&
{1 \over 4} \;
{{\cal Q}_{\!\tilde{d}} \over M_{\!\scriptscriptstyle \tilde{d}_m}^2} \left[
{\cal N}^{\!\scriptscriptstyle R}_{\!\scriptscriptstyle nmj} 
{\cal N}^{\!\scriptscriptstyle L^*}_{\!\scriptscriptstyle nmi} \;
{{M}_{\!\scriptscriptstyle \chi^0_n} \over m_{\!\scriptscriptstyle {d}_j}} \;
F_4\!\!\left({{M}_{\!\scriptscriptstyle \chi^0_{n}}^2 \over M_{\!\scriptscriptstyle \tilde{d}_m}^2} \right) \; 
+
{\cal N}^{\!\scriptscriptstyle L}_{\!\scriptscriptstyle nmj} 
{\cal N}^{\!\scriptscriptstyle L^*}_{\!\scriptscriptstyle nmi} 
 \;
F_2\!\!\left({{M}_{\!\scriptscriptstyle \chi^0_{n}}^2 
\over M_{\!\scriptscriptstyle \tilde{d}_m}^2} \right) \; 
\right]\;,\\
C^{\ssc \phi^-}_7 &=&
{1 \over M_{\!\scriptscriptstyle \tilde{\ell}_m}^2} \;
\widetilde{\cal C}^{\!\scriptscriptstyle  R}_{\!\scriptscriptstyle nmj} \,
\widetilde{\cal C}^{\!\scriptscriptstyle L^*}_{\!\scriptscriptstyle nmi} \;
{{m}_{\!\scriptscriptstyle u_n} \over m_{\!\scriptscriptstyle d_j}} \;
\left[  ({\cal Q}_{\!d} -{\cal Q}_{\!u}) \;
F_4\!\!\left({{m}_{\!\scriptscriptstyle u_n}^2 \over M_{\!\scriptscriptstyle \tilde{\ell}_m}^2} \right) 
- {\cal Q}_{\!u} \;
F_3\!\!\left({{m}_{\!\scriptscriptstyle u_n}^2 \over M_{\!\scriptscriptstyle \tilde{\ell}_m}^2} \right)  \right] \; 
\nonumber \\
&& +\;
{1 \over M_{\!\scriptscriptstyle \tilde{\ell}_m}^2} \;
\widetilde{\cal C}^{\!\scriptscriptstyle  L}_{\!\scriptscriptstyle nmj} \,
\widetilde{\cal C}^{\!\scriptscriptstyle L^*}_{\!\scriptscriptstyle nmi} \;
\left[  ({\cal Q}_{\!d} -{\cal Q}_{\!u}) \;
F_2\!\!\left({{m}_{\!\scriptscriptstyle u_n}^2 \over M_{\!\scriptscriptstyle \tilde{\ell}_m}^2} \right) 
- {\cal Q}_{\!u} \;
F_1\!\!\left({{m}_{\!\scriptscriptstyle u_n}^2 
\over M_{\!\scriptscriptstyle \tilde{\ell}_m}^2} \right)  \right]\;, \\ 
C^{\phi^{\mbox{\tiny 0}}}_7 &=&
{-{\cal Q}_{\!{d}} \over M_{\!\scriptscriptstyle S_m}^2} \left[
\widetilde{\cal N}_{\!\scriptscriptstyle nmj}^{\!\scriptscriptstyle R}\,
 \widetilde{\cal N}_{\!\scriptscriptstyle nmi}^{\!\scriptscriptstyle L^*} \;
{{m}_{\!\scriptscriptstyle d_n} \over m_{\!\scriptscriptstyle d_j}} \;
F_3\!\!\left({ m_{\!\scriptscriptstyle d_{n}}^2 \over 
M_{\!\scriptscriptstyle S_m}^2} \right)  \; 
+
\widetilde{\cal N}_{\!\scriptscriptstyle nmj}^{\!\scriptscriptstyle L}\,
 \widetilde{\cal N}_{\!\scriptscriptstyle nmi}^{\!\scriptscriptstyle L^*} \;
F_1\!\!\left({ m_{\!\scriptscriptstyle d_{n}}^2 \over 
M_{\!\scriptscriptstyle S_m}^2} \right)\right]\;.
\eea
Each term includes, implicitly,  summation over the $n$ fermion and $m$ scalar
mass eigenstates, except that the unphysical Goldstone modes are not to be
included. The latter contributions are incorporated into the gauge invariant
$C^{\ssc W}_7$ result. Contributions to $\widetilde{C}_7$ except for the case of
$C^{\ssc W}_7$ are obtained by replacement $L\leftrightarrow R$.
The $C^{\ssc W}_7$ term above is given in terms of explicit couplings,
rather than the effective coupling vertices. In accordance with the notation
of other terms, the vertices involved in the $C^{\ssc W}_7$ term above should be
two `$L$'-couplings the `$R$' counterparts of which vanish. Any term above of the
$LL$ type, {\it i.e.} with two `$L$'-couplings has, like $C^{\ssc W}_7$ term, a 
chirality flip from the $b$-quark external line. The rest of the terms, of the RL
type, have chirality flip inside the loop as illustrated also by the explicit
fermion mass ratio factor. Contributions from terms of RR type to $C_7$ coefficient
needs chirality flip from the $s$-quark external line and are neglected here, as
also in most of the literature. One can similarly map out the details for the 
chirality structure for the ${\wtl C}_7$ coefficient. We note, however, that the 
latter receives extra contributions from yet another class of diagrams with 
 $\l''_{ijk}$-couplings and the chirality flip on the $b$-quark. Such contributions 
are missing in the coefficient $C_7$, except for the ones suppressed with the
chirality flip from the $s$-quark. 
\bea
{\wtl C}^{\l''}_7 &=&
\frac{1}{4M_{\!\scriptscriptstyle \tilde{u}_m}^2}
\l^{''*}_{hnj}\l''_{kni}{\cal D}^{u*}_{h+3,m}{\cal D}^{u}_{k+3,m}
\left[-{\cal Q}_d F_1\left(\frac{m^2_{d_n}}{M_{\!\scriptscriptstyle \tilde{u}_m}^2}
\right) + {\cal Q}_{\!\tilde{u}}
F_2\left(\frac{m^2_{d_n}}{M_{\!\scriptscriptstyle \tilde{u}_m}^2}
\right)\right]\nonumber \\
&&+\;
\frac{1}{4M_{\!\scriptscriptstyle \tilde{d}_m}^2}
\l^{''*}_{nhj}\l''_{nki}{\cal D}^{d*}_{h+3,m}{\cal D}^{d}_{k+3,m}
\left[-{\cal Q}_u F_1\left(\frac{m^2_{u_n}}{M_{\!\scriptscriptstyle \tilde{d}_m}^2}
\right) + {\cal Q}_{\!\tilde{d}}
F_2\left(\frac{m^2_{u_n}}{M_{\!\scriptscriptstyle \tilde{d}_m}^2}
\right)\right]\;.
\eea
Similarly one can obtain the expressions for the $C_8$ and $\wtl{C}_8$ by introducing 
the colour factors at the relevant places. The loop-functions $F_{1-4}$
are the Inami-Lim functions, given as:
\bea
F_1(x) &=& 
\frac{1}{12 \,(1-x)^4} \, (2+3 \, x-6 \, x^2+x^3+6 \, x  \, \ln x) \;,
 \\
F_2(x) &=& \frac{1}{12 \, (1-x)^4} \, (1-6 \, x+3 \, x^2+2 \, x^3-6 \, x^2 \, \ln x) \; ,
 \\
F_3(x) &=&
\frac{1}{2(1-x)^3} \, (-3+4 \, x-x^2-2 \,  \ln x) \; ,\\
F_4(x) &=& \frac{1}{2(1-x)^3} \, (1-x^2+2 \, x \,\ln x) \; .
\eea
Below we write down the non-zero effective Wilson coefficients 
for the current-current operators.
\bea
C_2 &=& -\frac{G_F}{{\sqrt 2}}
V^{cs*}_{\!\mbox{\tiny CKM}}V^{cb}_{\!\mbox{\tiny CKM}}\;, \\
C_9& =& -\frac{1}{8M^2_{{\!\scriptscriptstyle S}_m}}
\l'_{i31}\l'^*_{j21} {\cal P}^s\;,  \\
C_{10}& =& -\frac{1}{8M^2_{{\!\scriptscriptstyle S}_m}}
\l'_{i32}\left[\l'^*_{j22} {\cal P}^s + y_s {\cal P}^{s'}\right]\;, \\
C_{11}& =& -\frac{1}{8M^2_{{\!\scriptscriptstyle S}_m}}
\l'^*_{j23}\left[\l'_{i33} {\cal P}^s + y_b {\cal P}^{s'}\right]\;, \\ 
\wtlc_1 &=& -\frac{1}{8m^2_{{\tilde d}_m}}
\l''_{2i2}\l^{''*}_{2j3}{\cal D}^d_{3+j,m}{\cal D}^{d*}_{3+i,m} 
= - \wtlc_2\;,\\
\wtlc_9& =& -\frac{1}{8M^2_{{\!\scriptscriptstyle S}_m}}
\l'_{i12}\l'^*_{j13} {\cal P}^s\;, \\
\wtlc_{10}& =& -\frac{1}{8M^2_{{\!\scriptscriptstyle S}_m}}
\l'^*_{j23}\left[\l'_{i22} {\cal P}^s + y_s {\cal P}^{s'}
\right]\;,  \\
\wtlc_{11}& =& -\frac{1}{8M^2_{{\!\scriptscriptstyle S}_m}}
\l'_{i32}\left[\l'^*_{j33} {\cal P}^s + y_b {\cal P}^{s'}\right]\;,  
\eea
\bea
\wtlc_{12}& =& \frac{-1} {8M^2_{{\!\scriptscriptstyle \tilde{\ell}_m}}}
\l'_{i12}\l'^*_{j13} 
{\cal D}^l_{2+i,m}{\cal D}^{l*}_{2+j,m}\;, \nonumber\\
&& -\;  \f{1}{8M_{\!\scriptscriptstyle \tilde{\ell}_m}^2}
y_s y_b {\cal D}^l_{2,m} {\cal D}^{l*}_{2,m}
V^{us*}_{\!\mbox{\tiny CKM}}V^{ub}_{\!\mbox{\tiny CKM}}\nonumber \\
&& -\; \f{1} {8M_{\!\scriptscriptstyle \tilde{\ell}_m}^2}
\l'^*_{j13}V^{us}_{\mbox{\tiny CKM}}y_s
{\cal D}^l_{2+i,m}{\cal D}^{l*}_{2,m} \;, \\
\wtlc_{13}& =& -\frac{1}{8M_{\!\scriptscriptstyle \tilde{\ell}_m}^2}
\l'_{i22}\l'^*_{j23} 
{\cal D}^l_{2+i,m}{\cal D}^{l*}_{2+j,m} \nonumber \\
&& -\;  \f{1}{8M_{\!\scriptscriptstyle \tilde{\ell}_m}^2}
y_s y_b {\cal D}^l_{2,m} {\cal D}^{l*}_{2,m}
V^{cs*}_{\!\mbox{\tiny CKM}}V^{cb}_{\!\mbox{\tiny CKM}}\nonumber \\
&& -\; \f{1}{8M_{\!\scriptscriptstyle \tilde{\ell}_m}^2}
\l'^*_{j23}V^{cs}_{\mbox{\tiny CKM}}y_s
{\cal D}^l_{2+i,m}{\cal D}^{l*}_{2,m} \;, \\
\wtlc_{14} &=& -\frac{1}{8m^2_{{\tilde d}_m}}
\l''_{1i2}\l^{''*}_{1j3}{\cal D}^d_{3+j,m}{\cal D}^{d\dagger}_{3+i,m} 
= -\wtlc_{15}\;, \\
\wtlc_{16} &=& -\frac{1}{8m^2_{{\tilde u}_m}}
\l''_{i12}\l^{''*}_{j13}{\cal D}^u_{3+j,m}{\cal D}^{u\dagger}_{3+i,m} 
=-\wtlc_{17}\;,
\eea
where,
\bea
{\cal P}^s &=& \f{1}{2}\left({\cal D}^s_{2+i,m} + i {\cal D}^s_{7+i,m}\right) \,\left(
{\cal D}^s_{2+j,m} + i{\cal D}^s_{7+j,m}\right)^{*}\;,  \\
{\cal P}^{s'} &=& \f{1}{2}\left({\cal D}^s_{2+i,m} + i {\cal D}^s_{7+i,m}\right) \,\left(
{\cal D}^s_{2,m} + i{\cal D}^s_{7,m}\right)^{*}\;.
\eea
Here, $y_s$ and $y_b$ denote the strange and the bottom Yukawas. Again the sum 
over repeated indices is assumed and the unphysical Goldstone modes are to be 
dropped from the sum. Note that the terms proportional to Yukawas are missing in the 
ref \cite{besmer}. This could lead to interesting contributions with the $\l'$ on one 
vertex and a SM Yukawa on the other vertex with a RPV mass mixing effect hidden 
within the ${\cal P}^{s'}$ expression. In the language of mass-insertion approximation, 
these kind of contributions involve a RPV mass-insertion ({\it e.g.} from a $\m_i$ or a 
$B_i$ coupling) along the scalar propagator.
\section{The Anomalous Dimension Matrix and RG Running}
The RG evolution from the scale $M_{\!\ssc W}$ to the scale appropriate for the decay 
dynamics ($m_b$) is described by eq.(\ref{rge}). This requires the knowledge of the 
anomalous dimension matrix $\g_{ij}$ (ADM). Derivation of $\g_{ij}$ presents many 
subtleties and was a main obstacle in the consistent calculation of the decay rate 
about a decade ago \cite{bsg_classics,cella}.
Below we mention the salient features of the calculation.
\begin{itemize}
\item 
Since QCD does not know the sign of $\g_5$ there is no mixing 
among the sets of ${\wtl {\cal Q}_i}$ and ${\cal Q}_i$ operators.
Thus the full basis of 28 operators splits into two invariant 
sub-spaces of 11 and 17 operators related by $L\leftrightarrow R$ inter-change.
\item 
An operator of a dimensionality $n$ can mix into the 
operators of dimensionality $\leq n$. Thus the current-current
four quark operators influence the RG evolution of (chromo)magnetic
Wilson coefficients, but not the other way around.
\item 
At the one-loop level current-current operators do not mix with
the (chromo)magnetic penguins. Thus, one has to evaluate this mixing 
at the two loop-level while still working at the leading log approximation.
Because of this, the ADM and hence the Wilson coefficients at the LL are found 
to be regularization scheme-dependent. Such a scheme-dependence
cancels with the corresponding scheme-dependence of the possible finite
one-loop (but $O(\a^0_s)$) contributions from certain four-quark operators
to the matrix elements of \bsg\ . Such contributions vanish in any four
dimensional as well as t-Hooft-Veltman regularization scheme but not
in naive dimensional regularization. Such a situation is taken care of
by expressing the ADM and the Wilson coefficients in a scheme independent
manner \cite{buras-scheme}.
\item 
Since one needs to perform a two-loop calculation, one must properly take 
into account the contributions from the so called evanescent operators (that vanish 
in the limit $D = 4$ dimension) while working in the $D\neq 4$ dimension.
\end{itemize}
With these remarks in mind one can formally write down the decay
amplitude as:
\be
A = C_7 \lag\, s\g\,|{\cal Q}_7|\,b\rag_{tree} + \widetilde{C}_7 
\;\lag s\g\,|\widetilde{{\cal Q}}_7|\,b\rag_{tree} +
\sum_{i=1}^{11}C_i \lag\,s\g\,|\cq_i|\,b\rag_{1-loop} +
\sum_{i=1}^{17}\wtlc_i\lag\,s\g\,|\cqt_i|\,b\rag_{1-loop}\;.
\ee
In order to obtain a decay rate that is regularization scheme-independent,
we must express the decay amplitude $A$ in terms of scheme-independent effective
coefficients. Decay amplitude $A$ is written as:
\be
A = C^{\mathrm{eff}}_7\; \lag\, s\g\,|{\cal Q}_7|\,b\rag_{tree}\; +\; 
\widetilde{C}^{\mathrm{eff}}_7 
\;\lag s\g\,|\widetilde{{\cal Q}}_7|\,b\rag_{tree}\;.
\ee
To derive the effective coefficients one makes the observation that in whatever 
scheme one chooses to work, the one-loop matrix elements of the current-current 
operators for the decay \bsg\ is always proportional to the tree-level matrix 
elements of the (chromo)magnetic operators. Thus, one can write
\bea
<s\g|{\cal Q}_i|b>_{\mathrm{1-loop}} & = & y_i<s\g|{\cal Q}_7|b>_{\mathrm{tree}}, \nonumber \\
<s\,\mathrm{gluon}|{\cal Q}_i|b>_{\mathrm{1-loop}} & = & 
z_i<s\, \mathrm{gluon}|{\cal Q}_8|b>_{\mathrm{tree}}\;,  \\
<s\g|{\widetilde {\cal{Q}}}_i|b>_{\mathrm{1-loop}} & = & 
\widetilde{y}_i<s\g|{\widetilde {\cal{Q}}}_7|b>_{\mathrm{tree}}\;,  \\
<s\,\mathrm{gluon}|{\widetilde {\cal{Q}}}_i|b>_{\mathrm{1-loop}} & = & 
\widetilde{z}_i<s\,\mathrm{gluon}|{\widetilde {\cal {Q}}}_8|b>_{\mathrm{tree}} \;.
\eea
In order to find out the proportionality factors $y_i, z_i$ etc., one
has to evaluate the finite contribution due to the insertion of certain
four-quark operators into the diagrams with a closed $b$ quark loop 
(in the NDR scheme). The only operators that could give finite contribution
are the ones with the chirality structure (LL)(RR) or (RR)(LL).
Within SM basis, these are $\cq_5,\cq_6$. With RPV you could also have
$\cqt_5,\cqt_6$ and $\cq_{11},\cqt_{11}$. The results for 
$\{y_i\}$, $\{z_i\}$, $\{\widetilde y_i\}$ and $\{\widetilde z_i\}$ are given as:
\be
\begin{array}{lcllcl}
y_i&=&\left\{\begin{array}{cl}-\frac{1}{3}&i=\cq_5\\
				-1&i=\cq_6,\:\cq_{11}\\
				0&\mbox{otherwise}
	\end{array}\right.
&z_i&=&\left\{\begin{array}{cl}1&i=\cq_5\\
				0&\mbox{otherwise}
	\end{array}\right.\\
	\\
\widetilde y_i&=&\left\{\begin{array}{cl}-\frac{1}{3}&i=\cqt_{5}\\
				-1&i=\cqt_{11},\:\cqt_{6}\\
				0&\mbox{otherwise}
	\end{array}\right.
&\widetilde z_i&=&\left\{\begin{array}{cl}1&i=\cqt_{5}\\
				0&\mbox{otherwise}
	\end{array}\right.
\end{array}
\ee
With this one can write down the scheme-independent Wilson coefficients as:
\begin{eqnarray}
C_7^{\mathrm{eff}}(\mu)&=&C_7(\mu)+\sum_{i=1}^{11} y_i \; C_i(\mu)\;,\nonumber \\
C_8^{\mathrm{eff}}(\mu)&=&C_8(\mu)+\sum_{i=1}^{11} z_i  \; C_i(\mu)\;, \nonumber \\
\widetilde C_7^{\mathrm{eff}}(\mu)&=&\widetilde C_7(\mu)+\sum_{i=1}^{17}\widetilde y_i 
 \; \widetilde C_i(\mu)\;,\nonumber \\
\widetilde C _8^{\mathrm{eff}}(\mu)&=&\widetilde C_8(\mu)+\sum_{i=1}^{17}\widetilde z_i 
 \; \widetilde C_i(\mu)\;,\nonumber \\
C_{\cq_i,\cqt_i}^{\mathrm{eff}}(\mu)&=&C_{\cq_i,\cqt_i}(\mu)\;.
\end{eqnarray}
The index $i$ above runs over only the current-current operators.

The RG evolution of the  effective coefficients is determined by the
 scheme-independent effective ADM. The RG equations are given as:
\be
\frac{d}{\ln\mu}C_k^{\mathrm{eff}}(\mu)=\frac{\alpha_s}{4\pi}\gamma_{jk}^
{\mathrm{eff}}  \; C_j^{\mathrm{eff}}(\mu)\;.
\ee
Here $C_k^{\mathrm{eff}}$ are the components of a column vector 
$\vec{C}^{\mathrm{eff}}$ which contains coefficients $C_k^{\mathrm{eff}}$ as well 
as ${\wtl C}_k^{\mathrm{eff}}$.  Since QCD does not know the sign of $\g_5$, these 
two sets of coefficients do not mix. The effective ADM at leading log is obtained as:
\be
{\hat \g}^{(0)\mathrm{eff}} =
\bmat{cc}
\g^{\mathrm{eff}}_{\ssc \!L} & 0\\
0& \g^{\mathrm{eff}}_{\ssc \!R}\\
\emat
\;.\ee
$\g^{\mathrm{eff}}_{\ssc \!L}$ represents QCD mixing of 11 operators whose 
chirality structure is similar to those of SM operators. 
$\g^{\mathrm{eff}}_{\ssc \!R}$ represents the mixing of 17 operators whose chirality
structure is obtained by $L\leftrightarrow R$ replacement with SM like operators.
We present the explicit matrices in the appendix \ref{gamma}.

In general, the solution of the RGE for the Wilson-coefficients is given by
\be
\vec
C^{\mathrm{eff}}(\mu)=V\left[\left(\frac{\alpha_s(M _{\!\ssc W})}{\alpha_s(\mu)}\right)^{\vec
\gamma^{\mathrm{eff}}_{\!\ssc D}/2\beta_0} \right]_DV^{-1}\; 
\vec C^{\mathrm{eff}}(M_{\!\ssc W})\;,
\ee
where $V$ diagonalizes $(\gamma^{\mathrm{eff}})^T$
\be
\gamma^{\mathrm{eff}}_{\!\ssc D}=V^{-1}(\gamma^{\mathrm{eff}})^{T} V\;.
\ee
$\beta_0=23/3$ is the one-loop beta-function and 
$\vec\gamma^{\mathrm{eff}}_{\!\ssc D}$ is the 
vector containing the eigenvalues of $\gamma^{\mathrm{eff}}$. 
In our case
\be
\vec\gamma^{\mathrm{eff}}_{\!\ssc D}=\begin{array}[t]{ccccccc}(-16&-16&-16&-16&-16&-16&-16\\
-16&-8&-8&-8&-8&4&4\\
4&4&\frac{28}{3}&\frac{28}{3}&\frac{32}{3}&\frac{32}{3}&2.233\\
2.233&6.266&6.266&-13.791&-13.791&-6.486&-6.486)\;.\end{array}
\ee
With $\alpha_s(M_{\!\ssc Z})=0.121$ and $\mu=m_b=4.2$ GeV, the coefficients
$C_7^{\mathrm{eff}}(m_b)$ and $\widetilde C_7^{\mathrm{eff}}(m_b)$ are given as: 
\begin{eqnarray}
\label{c_mb}
C_7^{\mathrm{eff}}(m_b)&=& 
-0.351 \; C_{2}^{\mathrm{eff}}(M_{\!\ssc W})
+0.665 \;  C_{7}^{\mathrm{eff}}(M_{\!\ssc W})
+0.093 \; C_{8}^{\mathrm{eff}} (M_{\!\ssc W})
-0.198 \; C_{9}^{\mathrm{eff}}(M_{\!\ssc W}) \nonumber \\
&&-0.198 \;  C_{10}^{\mathrm{eff}} (M_{\!\ssc W})
-0.178 \;  C_{11}^{\mathrm{eff}}(M_{\!\ssc W})\;, \nonumber \\
[0.5cm]
{\wtl C}_7^{\mathrm{eff}}(m_b)&=&
0.381 \;  {\wtl C}_{1}^{\mathrm{eff}}(M_{\!\ssc W})
+0.665  \; {\wtl C}_{7}^{\mathrm{eff}}(M_{\!\ssc W})
+0.093  \; {\wtl C}_{8}^{\mathrm{eff}}(M_{\!\ssc W})
-0.198  \; {\wtl C}_{9}^{\mathrm{eff}}(M_{\!\ssc W}) \nonumber \\
&&-0.198 \;  {\wtl C}_{10}^{\mathrm{eff}}(M_{\!\ssc W})
-0.178  \; {\wtl C}_{11}^{\mathrm{eff}}(M_{\!\ssc W})
+0.510 \;  {\wtl C}_{12}^{\mathrm{eff}}(M_{\!\ssc W})
+0.510 \; {\wtl C}_{13}^{\mathrm{eff}}(M_{\!\ssc W})\nonumber \\
&&+0.381 \; {\wtl C}_{14}^{\mathrm{eff}}(M_{\!\ssc W})
-0.213 \; {\wtl C}_{16}^{\mathrm{eff}}(M_{\!\ssc W})\;.
\end{eqnarray}
The numbers multiplying the different Wilson coefficients are all of the same size, 
hence no term can be neglected {\it a priori}. 

Finally, the branching fraction for $Br (b \rightarrow s + \g )$  is expressed through 
the semi-leptonic decay $b \rightarrow u|c e{\bar \nu}$ so that the large bottom 
mass dependence $( \sim m^5_b)$ and uncertainties in CKM elements cancel out.
\be 
Br (b \rightarrow s + \g) = \frac{\Gamma (b \rightarrow s + \gamma )}
{\Gamma (b \rightarrow u|c \,e\,{\bar \nu_e})}
Br_{\mathrm{exp}} (b \rightarrow u|c \,e\,{\bar \nu_e})\;,
\ee
where $Br_{\mathrm{exp}} (b \rightarrow u|c\,e\,{\bar \nu_e}) = 10.5 \%$.
Here the presence of RPV couplings offers new decay channels and new 
contributions to the semi-leptonic decay rate. In principle, lepton flavor violating couplings demand a summation over the three (anti)neutrino mass eigenstates 
(assuming the neutralinos are heavy) before squaring the amplitude. 
In considering the $\l'$-coupling contributions,
the dominating neutrino mass terms would be at one-loop level. As all the neutrino
states have close to zero masses, and very small admissible mixings with the
gauginos and the Higgsinos, we neglect these effects here and stick to effective
massless electroweak neutrinos. The summation is then taken simply over the three
families. Under the same spirit, we neglect here the effects from the lepton 
flavor violating couplings that could change the electroweak character of physical
electron. The partial widths in the expression above are given, at the LL order
with $m^2_u/m^2_b$ set to zero, as
\bea
\Gamma (b \rightarrow s \g) & = & \f{\a m^5_b}{64 \pi^4} (|C^{\rm eff}_7 (\m_b)|^2 + 
|\widetilde{C}^{\rm eff}_7(\m_b)|^2) \;,
\nonumber \\
\Gamma (b \rightarrow u|c\; e\; \bar{\n}) & = & \f{m_b^5}{192\pi^3}
\frac{1}{32}
\left\{f(\e)\left[|2A + C_2|^2 + B_2\right] 
 +\; |B_1|^2 + |C_1|^2\right\} \;,\nonumber\\
f(\e) & = & 1 - 8 \e^2 + 8 \e^6 - \e^8 - 24\e^4 \mathrm{log}\e \;,  \nonumber \\
A & = & 2\sqrt{2}G_{\!\ssc F}V^{ts}_{\!\mbox {\tiny CKM}}\;,\nonumber\\
B_r & = & \sum_{i=1}^{3}\frac{1}
{M_{\!\scriptscriptstyle \tilde{\ell}_l}^2}\left[
{\l_{ij1}\l^{'*}_{mn3}}
D^{l*}_{2+m,l}D^l_{2+j,l} \right]V^{rn}_{\!\mbox {\tiny CKM}} \;
+\; \left[{h_e \l^{'*}_{mn3}D^{l*}_{2+m,l}D^l_{2,l}}\right]
V^{rn}_{\!\mbox {\tiny CKM}} \;\;\;\;r = 1,2 \;\;,\nonumber \\
C_r & = & - \sum_{i=1}^{3}\frac{\l'_{i3k}\l^{'*}_{1mn}}
{M_{\!\scriptscriptstyle \tilde{d}_l}^2}
D^{d*}_{3+k,l}D^d_{3+n,l}V^{rm}_{\!\mbox{\tiny CKM}}\;. \nonumber 
\eea
Here $\e  =  m_c/m_b$. The above expression for the semi-leptonic decay rate
is the most general one that also includes the contribution from bilinear-trilinear 
combination of $R$-parity violating parameters (the term $\propto h_e$) 
which has been generally missed in the earlier studies.
Such contributions have been shown to play a very 
important role as elaborated in detail in our parallel report on \bsg\ \cite{017} and for the
case of quark dipole moments in \cite{kong-keum}.
\section{Numerical Results on the Trilinear Couplings}
Our presentation of the numerical results in this paper restricts to the effects of the combinations of trilinear couplings only. We present some details of the numerical 
results on the phenomenologically interesting $\l'$-couplings. The $\l''$-couplings
are found not to give interesting results --- a feature we will comment on at the
end of the section. To set the stage for the more detailed discussion on our numerical 
results, we first recall the salient features of this particular decay within the MSSM.
Within MSSM, the story is already quite complicated. In the limit of exact SUSY
there is no contribution to the decay rate due to the cancellations among particles 
and sparticles \cite{ferrara}. With SUSY broken, there could be gluino, chargino, 
charged Higgs and neutralino contributions depending on what particle is exchanged 
in the loop. Chirality flip can be induced by mass-insertion on the internal or the 
external fermion line. The squark mass matrices are not necessarily flavor
diagonal and the flavor violation is introduced as family off-diagonal mass
insertion in the loop. The couplings of quarks and squarks to the neutral
gauge bosons and gauginos are flavor diagonal. The more exact treatment, 
as adopted here, is to use the mass eigenstates for the particles running in the
loop and hence have the flavor violation effects absorbed into the effective
couplings which contain the corresponding elements of the squark diagonalizing 
matrices. With the inclusion of RPV couplings, there are a lot more flavor
violating couplings, and more new types of admissible diagrams. These are 
described in our mass eigenstate expressions most naturally by simply enriching 
the effective couplings.

The significance of the MSSM contribution to the decay rate is  highly dependent on 
the choice of values for the background parameters of the model.
Over and above the delicate issue of cancellations among various contributions,
it is also a function of `how much' and `where' does one introduces the
flavor violation. The approaches in the literature can be broadly classified into two 
categories. (a) The so called constrained MSSM (cMSSM) where one starts with the 
universal boundary conditions at the high scale and derives the weak-scale spectrum 
by RG evolution. CKM angles are then the sole source of flavor violation\cite{mfv}.
In such a scenario, the dominant contributions are due to the chargino and charged 
Higgs loops. Gluino and neutralino couplings being flavor diagonal, rely on flavor 
violating mass mixing terms in the down squark mass matrix generated from RG 
evolution. Such mixings within cMSSM are typically very small to contribute to the 
decay rate. Gluino contributions suffer from hard GIM cancellations. While neutralino
contributions can escape GIM cancellations through Yukawa couplings but cannot 
compete with chargino contribution that are enhanced by heavy top in the 
loop\cite{baer}. 
(b) Another end of the spectrum is simply the unconstrained scenario. Here, the idea is 
to assume arbitrary structure for the squark mass matrices and derive the weak-scale 
bounds on the off-diagonal flavor violating entries by requiring to reproduce
the decay rate within the experimental uncertainties \cite{greub-bes}. We adopt the 
philosophy of unconstrained scenario studies. However, we will switch-off the 
$R$-parity conserving flavor violations, except that from the CKM matrix to focus on 
the effect of the RPV parts. In fact, we will focus on a pair of $\l'$-couplings (or 
$\l''$-couplings) at a time while the other flavor violating parameters that are
otherwise not theoretically constrained will be switched off. Note that the flavor 
violations originating from $\l'$-couplings are essentially of supersymmetric origin 
though SUSY breaking, and phenomenologically viable, scalar and gaugino masses 
are needed to avoid the intrinsic SUSY cancellations.

\begin{figure}
\includegraphics[scale=1.2]{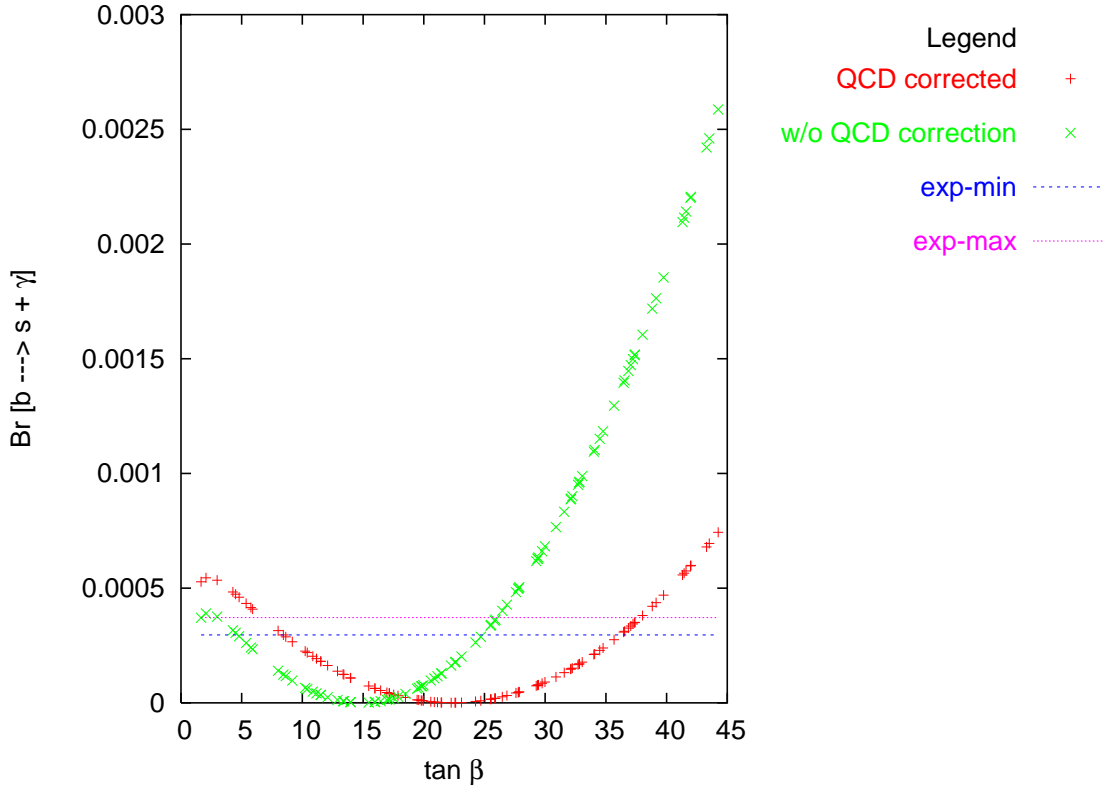}
\caption{\label{mssmn-rate-} 
Branching ratio as a function of $\tan\!\b$ in MSSM with
$A_t$ positive and $\m$  negative. The `+' sign is for the QCD corrected rate
and the `$\times$' sign for the rate without QCD corrections. Two horizontal
lines in this Fig. and all the following figures for branching fraction is the 
experimental uncertainty at $1 \s$. See text for the values of various parameters.}
\end{figure}

\begin{figure}
\includegraphics[scale=1.2]{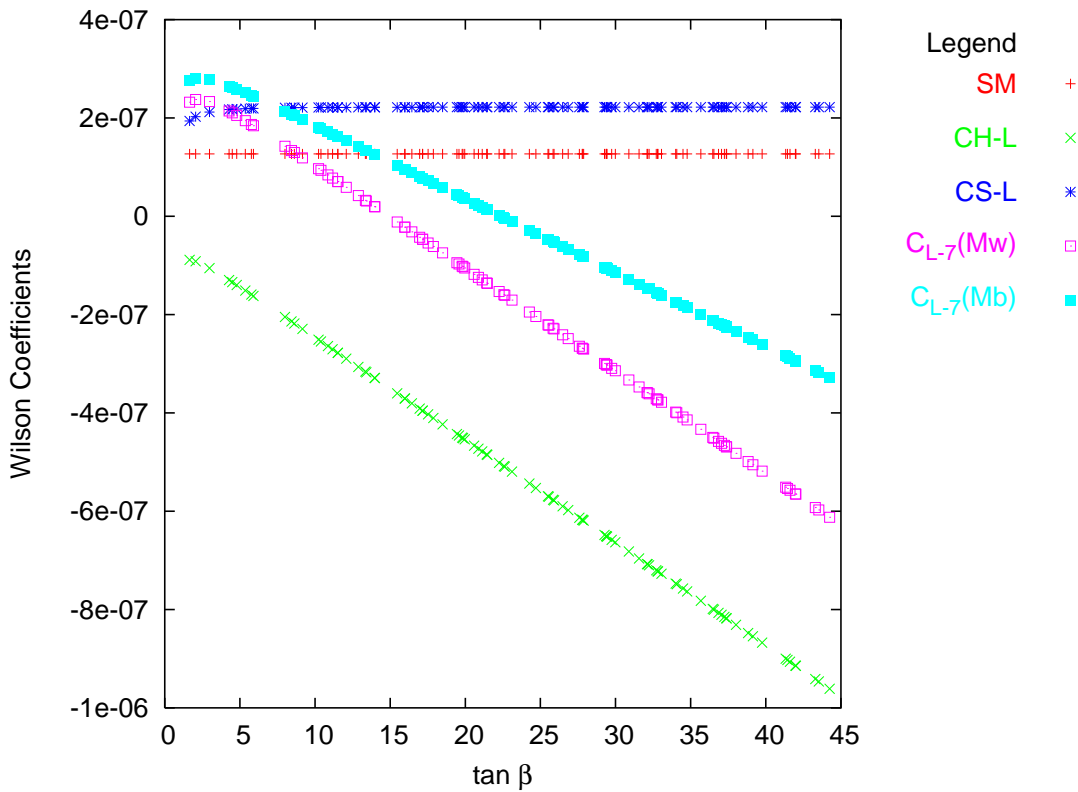}
\caption{\label{mssm-wil} 
Various contributions to Wilson coefficients versus $\tan\! \b$. SM 
(line with symbol `+') and charged Higgs (denoted as CS-L, line with `*' sign) are of
 the same sign (positive) whereas the chargino contribution (denoted as CH-L, line 
with 'x' sign) has a strong $\tan \!\b$ dependence and is of opposite sign (negative) 
because $A_t \m < 0$.}
\end{figure}

\begin{figure}
\includegraphics[scale=1.2]{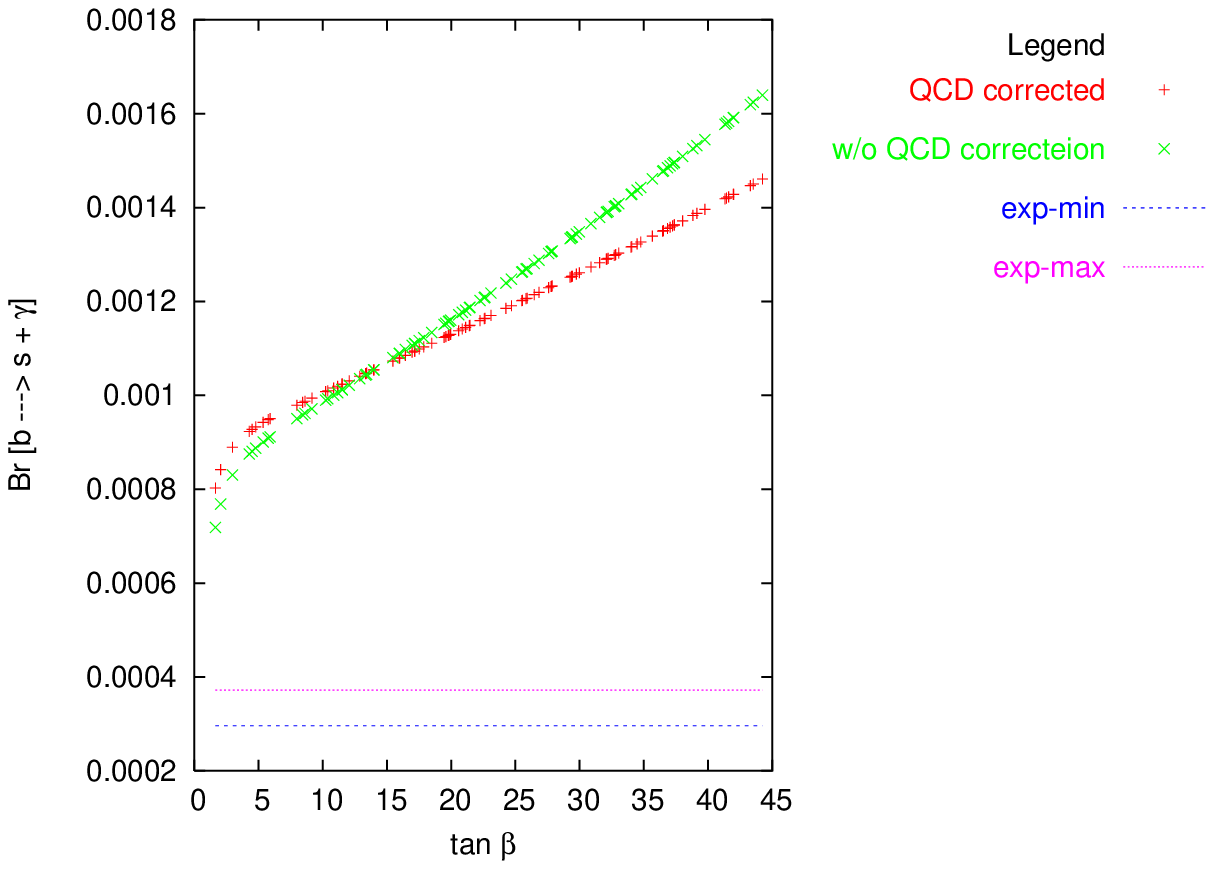}
\caption{\label{mssmn-rate} 
Same as Fig.\ref{mssmn-rate-}, but with sign of $\m$ positive.}
\end{figure}

Before our analysis of RPV contributions, we have first checked our code for the case
of MSSM in order to reproduce the trends and features obtained in the previous
works. Here, the matching can be done at a qualitative level only. Most of the 
detailed numerical results in the literature for the case of MSSM are worked out within 
the cMSSM scenario. The low energy spectrum obtained by RG evolution and a 
particular SUSY breaking boundary condition contains correlations among various 
parameters and there are RG induced flavor violations mixed with that from the CKM 
matrix. With this limitations, our code does reproduce qualitatively, trends and features
of various contributions obtained by others \cite{mssm,baer,carena}. We give an illustration
of our MSSM results in Figs.\ref{mssmn-rate-},\ref{mssm-wil} and \ref{mssmn-rate}. For a scenario 
characterized by squarks around 300 GeV, sleptons around 150 GeV, down-type Higgs 
and $A$ parameter 300 GeV, (up-type Higgs mass and $B_{\ssc 0}$ parameter 
determined from potential minimization condition), gaugino $M_2 = 200$ GeV 
(assuming the relationship $M_1 = 0.5 M_2, M_3 = 3.5 M_2$) and $\m = -300$ GeV 
for Fig.\ref{mssmn-rate-} and 300 GeV for Fig.\ref{mssmn-rate} (note that the 
MSSM parameter $\m$ is denoted by $\m_{\ssc 0}$ in our model notation),
we plot branching fraction versus $\tan\! \b$. As is well known, the charged Higgs contribution is of the same sign as the SM whereas the sign of chargino contribution 
depends on the sign of the product $A_t \m$ (see Fig. \ref{mssm-wil} where we have 
plotted various contributions to the Wilson coefficients). In the large $\tan\!\b$ region,
the decay rate is completely dictated by the chargino contributions  and hence the 
enhancement or the suppression of the rate actually depends on whether $A_t\m$ is 
positive or negative, respectively. Within the framework of 
mSUGRA (see ref.\cite{carena}), choosing the gaugino masses to be positive, 
forces $A_t$ to be negative at the weak scale, and hence one requires a positive 
$\m$ for the rate at large $\tan\! \b$ to fall within the experimental limits. 
In Fig.\ref{mssmn-rate-},\ref{mssmn-rate} we plot the decay rate for 
negative and positive $\m$ respectively with $A_t$ fixed to be positive. One can see 
in the figures that the decay rate does fall initially  with $\tan\! \b$ and then 
rises in case of negative $\m$ but for positive $\m$, due to constructive interference 
between chargino and charged-Higgs the rate rises with $\tan\! \b$.
This can be easily compared with the Fig.2 of ref.\cite{carena}, for example. 

While investigating the influence of RPV we kept the spectrum similar to the one 
selected for the case of MSSM with sign of $\m$ negative and $\tan\!\b = 37$. 
Note that in the large $\tan\!\b$ region QCD corrections drastically suppress the rate 
at the scale $m_b$ (see for instance Fig.\ref{mssmn-rate-}). This is expected 
because, in the large $\tan\!\b$ region, the Wilson coefficient $C_7$ is negative due 
to large negative chargino contribution and at the scale $m_b$ after the
QCD correction there is a destructive interference between $C_2$ and $C_7$. 
The choice of spectrum is guided by requiring that in the MSSM limit we do obtain branching fraction well within experimental limits. Now the purpose is to investigate 
the behaviour of branching fraction once we turn on the knob of $R$-parity violation.
As we will see, both, enhancement as well as suppression is possible depending 
on which operator(s) are playing the major role. Both, the direct contribution, 
through the magnetic penguin, and the effect induced by QCD renormalization,
through the various four-quark operators are significant.

Due to the large number of RPV parameters (48 just from the superpotential, with
more from the soft SUSY breaking sector), without further knowledge of their plausible
range of values it is impossible to extract any useful information on a 
phenomenological study, with all of them playing a role at the same time. Hence,
the sensible thing to do at this stage is to focus on a minimal number at a time
and study their possible impact. For the case at hand, a pair of parameters is taken
at a time, while all the others are switched off. We discuss first the $\l'$-couplings.
There are two possible combinations in which two non-zero $\l'$ can give a finite 
contribution. These are  (A) $\l'_{i3j}\l^{'*}_{h2k}$ and (B) $\l^{'*}_{ij3}\l^{'}_{hk2}$. 
We illustrate their effects in more detail below.

We will compare below the probable bounds we obtained with the similar bounds 
on the relevant combination of $\l'$-couplings in the literature. We have, first, to 
bring to the readers attention that the existing bounds in the literature are 
typically obtained assuming a sparticle spectrum of around 100 GeV.
However, such a low lying spectrum is very dangerous for \bsg\ 
\footnote{ To be exact, the top squarks contribute directly through the chargino
diagram without the CKM suppression and have to be heavier. The bottom and strange
squarks contribute, if their flavor violating mixings are nonzero. They are 
hence likely to be required to be heavier too. Otherwise, we can actually live with
only heavier stops but a light sparticle here. Such a split in sparticle
spectrum is considered unlikely as it goes in the opposite direction as predictions
from most available SUSY breaking models.} 
and quite unrealistic at least for the squarks. In consideration of that, and for a 
better comparison with the available \bsg\ result for the $R$-conserving 
contributions (of the MSSM), we stick to a slightly heavier spectrum of 300 GeV 
squarks. Hence, the existing bounds must be rescaled by a factor of three for a better 
comparison with our bounds. Moreover, for some the combinations that contribute 
to \bsg\ there do not exist any direct bound. We have then made use of the best 
bounds available on the individual parameters to obtain a bound on such a combination.
\subsubsection{Case A: $\l'_{i3j}\l^{\prime *}_{h2k}$}
If the indices $i,j,h,k$ are un-constrained, one can form 81 combinations of
two $\l'$-couplings that can in principle contribute. However, in the limit there is no 
flavor off-diagonal mixing in the (s)neutrino mass 
matrix, the combination with $i\neq h$ will not contribute. The case $j\neq k$
requires extra source of flavor off-diagonal squark mixing to contribute; hence
is also not of interest here. So we shall confine ourselves to the case $i=h$ and $j=k$
and hence study the impact of only nine (9)  combinations, picking one of them
to be non-zero at a given time.  Note that so long as one sticks to the 
phenomenologically required suppression in all the flavor off-diagonal mass
mixings from SUSY, the RPV contributions which require such mixings would be 
doubly suppressed and hence highly unlikely to be of any numerical significance.
For the 
magnetic penguin operators, the contribution is of LL type. 

The above combination will lead to a $b_{\!\ssc R} \rightarrow s_{\!\ssc L}$ 
transition with $m_b$ mass-insertion on external line. The contribution can come 
from chromo(magnetic) penguins (only sneutrino and neutrino loops are possible) 
or from the current-current operators $\cq_{9,10,11}$. Sneutrino loops being proportional
to the inverse of light slepton mass, are dominant compared to the neutrino loops which are 
suppressed by heavier squark mass.
The important feature here 
is that the RPV contributions interfere with the SM and the MSSM contributions. 
Of the possible nine combinations, we will  discuss three 
representative combinations with best bounds. 
The other six combinations 
follow similar pattern, as their contributions  have a similar 
structure. 
\par
\begin{figure}
\includegraphics[scale=1.2]{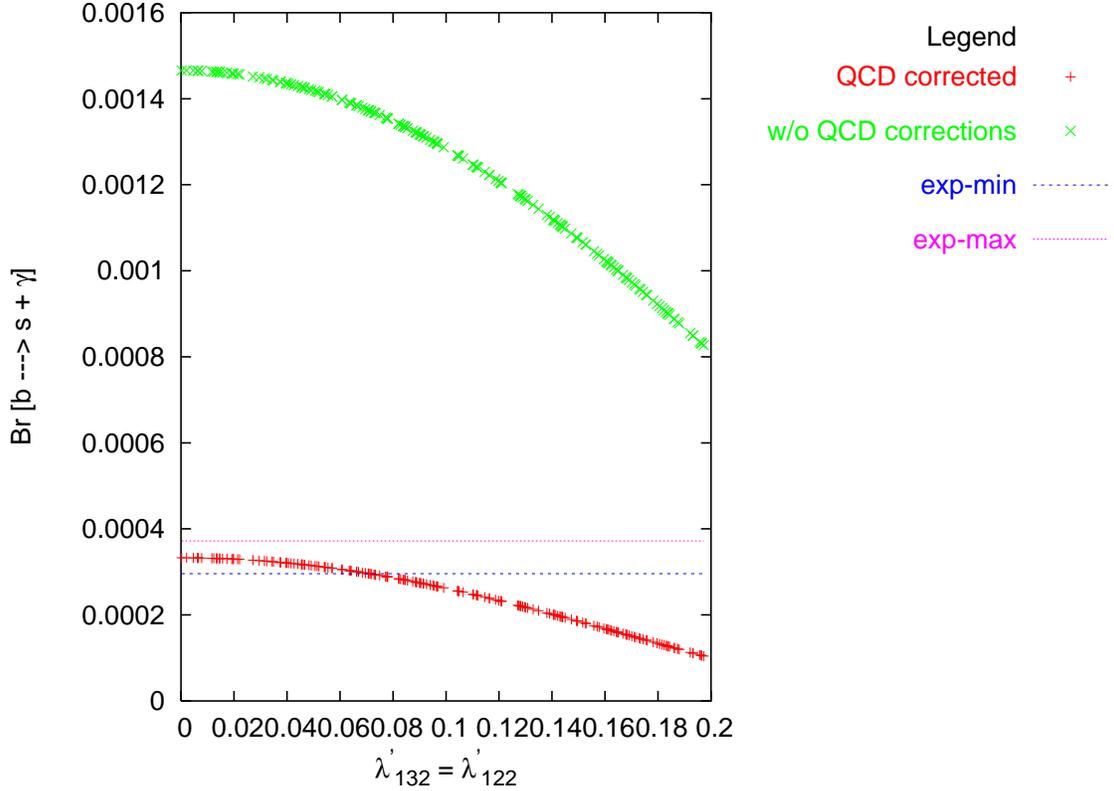}
\caption{\label{tt-132-122} Branching fraction as a function of
$\l'_{132}\;(=\l'_{122})$.}
\end{figure}

\begin{figure}
\includegraphics[scale=1.2]{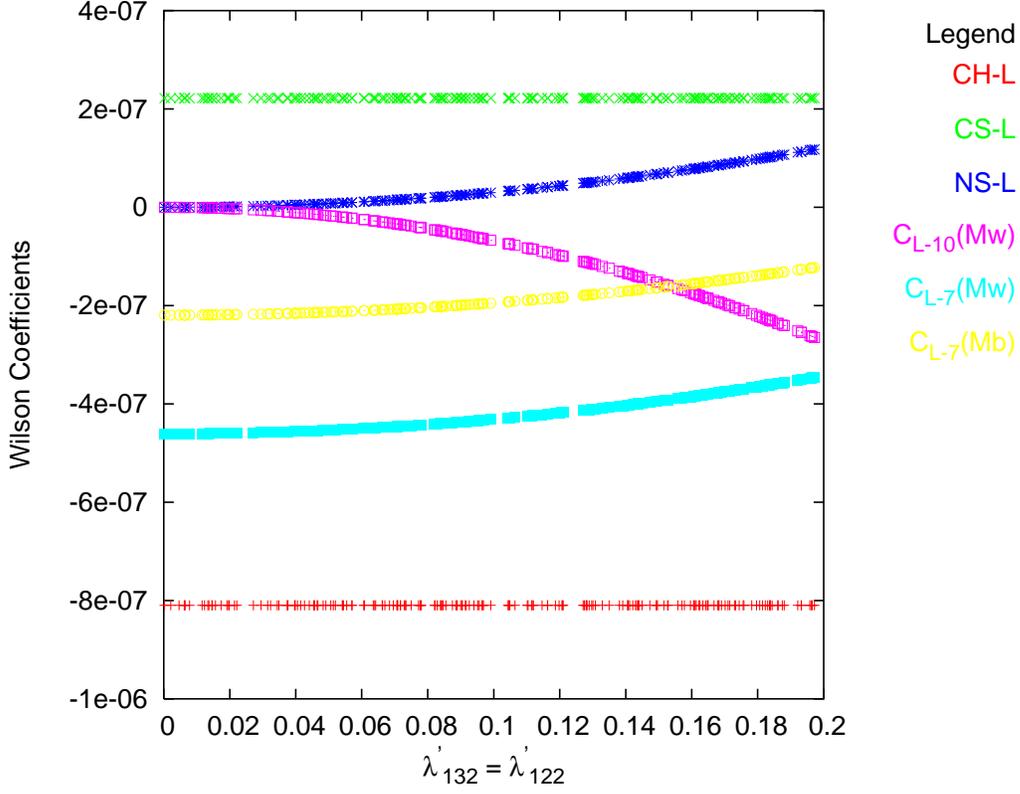}
\caption{\label{ttw-132-122} Wilson coefficients as a function of
$\l'_{132}\;(=\l'_{122})$. CH,CS,NS stand for the chargino, charged-Higgs
 and sneutrino contributions, respectively. The letter `L' indicates that these 
contributions go into $C_7$. Similar notation would apply to the following figures. 
The letter `R'(L) would indicate that the contribution goes into coefficients 
with(without) tilde. }
\end{figure}

 In our numerical analysis, we take real and equal values for two $\l'$-couplings
at a time. As discussed above and illustrated by the analytical formulae, the contributions
depend on the complex product. The strategy here then simplifies the results to 
one-parameter scenario for easy presentation, without compromising the physics.
Consider the influence of the combinations $|\l'_{13k}\l'_{12k}|$. For $k=1,2$ the 
physics is identical because the relevant Wilson coefficients ($C_{9,10}$) have 
identical structure and so we shall discuss the combination  $|\l'_{132}\l'_{122}|$.
In Fig.\ref{tt-132-122} we have plotted the branching ratio against 
$\l'_{132}\;(=\l'_{122})$. Interestingly, the rate falls below the experimental limit 
(at $\l'_{132}=\l'_{122} = 0.07$) with increasing $\l'$ value, reaches a minimum 
and then rises again (the rise is not shown in the fig.).  To understand this behavior 
we have plotted all the relevant Wilson coefficients against $\l'$ values in 
Fig.\ref{ttw-132-122}. It can be clearly seen that $C_7 (M _{\!\ssc W})$ is falling 
with increasing  $\l'$ value. This is because of destructive interference between the 
negative $R$-parity conserving contribution (sum of W-boson, charged Higgs and Chargino loop)
and the positive sneutrino loop contribution, as seen in the
figure. At the scale $m_b$ after QCD running there is again destructive interference 
between the dominant tree level contribution from $C_{10}$ and $C_7$ 
[see eq. (\ref{c_mb})], because of which $C_7 (m_b)$ further falls with 
increasing  $\l'$ value. So because of these destructive interference, the branching 
ratio is falling below the experimental limit and we obtain a bound of
$|\l'_{132}\l'_{122}|< 4.9\times 10^{-3}$, to be compared with a rescaled 
existing bound of $3\times 10^{-2}$.
\begin{figure}
\includegraphics[scale=1.2]{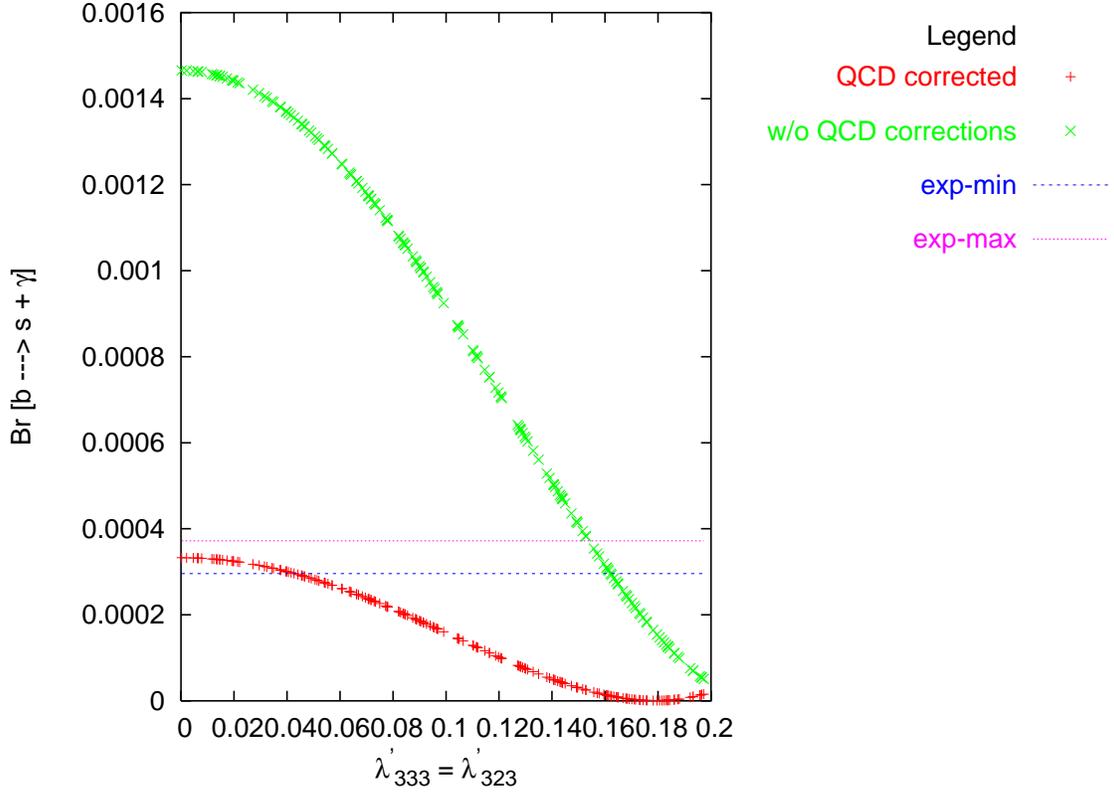}
\caption{\label{tt-333-323}   Branching fraction as a function
of $\l'_{333}\;(=\l'_{323})$.}
\end{figure}
\begin{figure}
\includegraphics[scale=1.2]{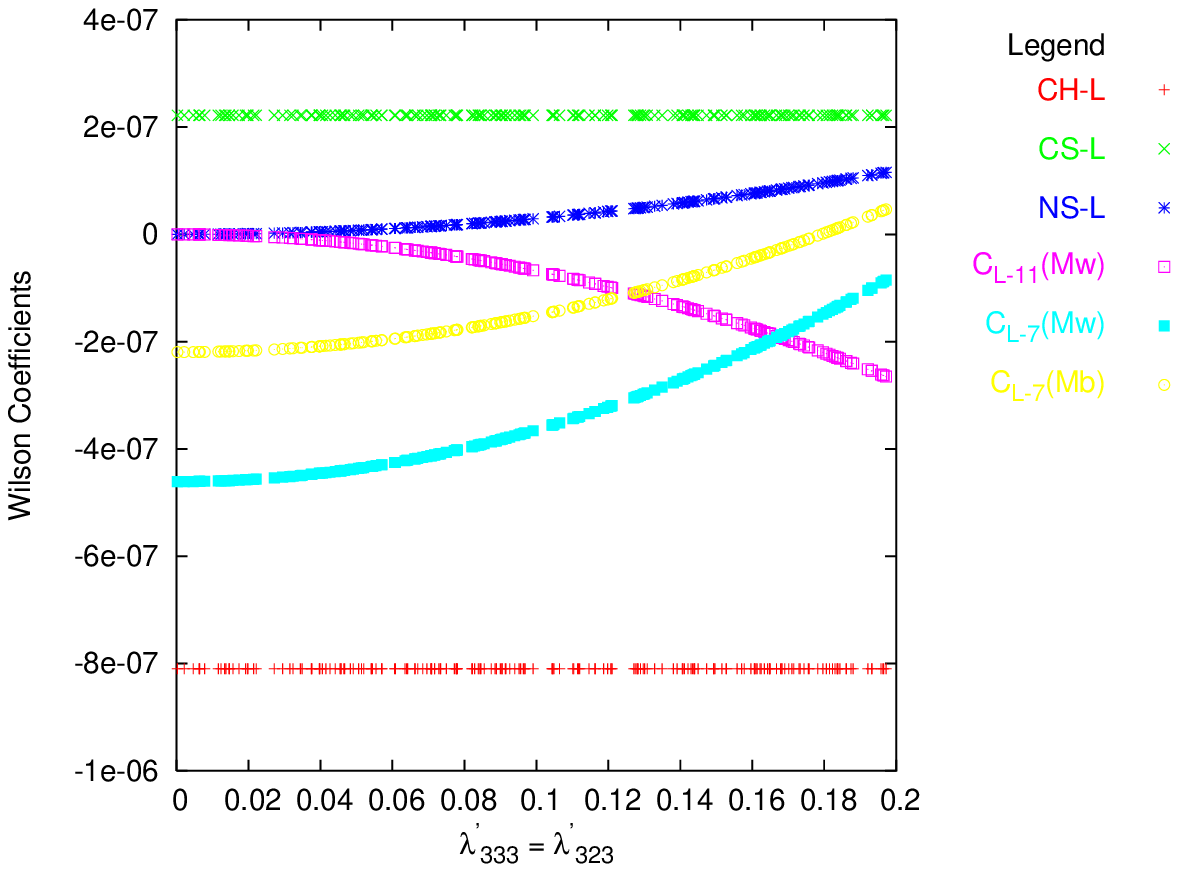}
\caption{\label{ttw-133-123}   Relevant Wilson coefficients as a function
of $\l'_{333}\;(=\l'_{323})$.}
\end{figure}
In Fig.\ref{tt-333-323}, we have plotted the branching ratio against 
$\l'_{333} \;(=\l'_{323})$. Unlike the above case, here the relevant current-current 
Wilson coefficient $C_{11}$ has a more important role to play. Not only does 
the scheme-independent  effective coefficient $C_{11}^{\rm eff}$ influence 
$C_7 (m_b)$ through RG running, it also contributes to $C_7^{\rm eff}$ directly. 
The effect cancels with the $R$-parity conserving contributions and hence  
$C_7^{\rm eff} (M_{\!\ssc W})$ is seen to reduce with increasing $\l'$ value. 
Even at the scale $m_b$, after QCD running there is a destructive interference 
between the $R$-parity conserving and $R$-parity violating parts and hence 
$C_7^{\rm eff}$ at $m_b$ further
reduces resulting in the branching fraction that falls with the RPV coupling. The fall 
is much stronger compared to the previous case. We obtain a bound of 
$|\l'_{333}\l'_{323}| < 1.6 \times 10^{-3}$ to be compared with the rescaled 
existing bound of $(1-3)\times 10^{-2}$ \cite{gautam}.
\subsubsection{Case B: $\l^{'*}_{ij3}\l'_{hk2}$}
Such a combination leads to a $b_{\!\ssc L}\rightarrow s_{\!\ssc R}$  transition
for $m_b$ mass insertion on the external line. This would mean that the $b$ and 
$s$ quark at the effective vertices would be right handed. It contributes to the Wilson 
coefficients of several operators. These include the  $\wtl{C}_{7,8}$ coefficients of
the {(chromo)magnetic penguins}, via charged lepton, charged slepton, neutrino and 
sneutrino loop. Interestingly, in the SM and MSSM, this coefficient does not receive 
much contribution. The above combination can also contribute to Wilson 
coefficients of the {current-current operators}, $\cqt_9$ to $\cqt_{13}$. These 
current-current contributions are mostly the dominant ones but not always as we will soon see. 
\par
\begin{figure}
\includegraphics[scale=1.2]{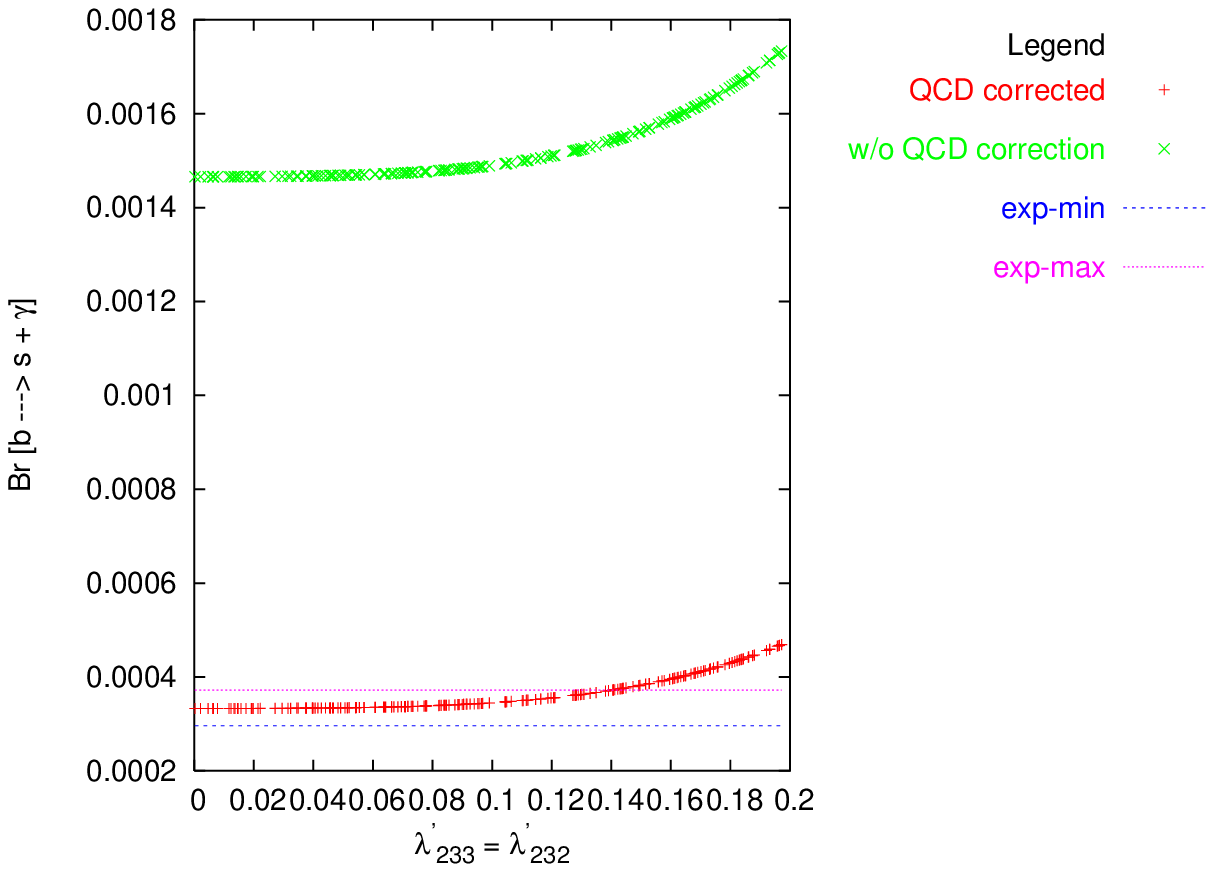}
\caption{\label{tt-233-232} 
Branching ratio as a function of $\l'_{233} \;( = \l'_{232})$. }
\end{figure}

	As discussed at length in case (A), we must require $i=h$ and $j=k$ and 
hence case (B) then effectively consists of the combinations $\l^{\prime *}_{ij3}\l'_{ij2}$.
Consider Fig.\ref{tt-233-232} where we have plotted the branching fraction against	
non-vanishing $\l'_{233}\;( = \l'_{232})$. We get a bound 
$ |\l'_{233}  \l'_{232}|< 1.9 \times 10^{-2}$,
to be compared with the rescaled existing bound of $4.4 \times 10^{-1}$ \cite{gautam}.
\begin{figure}
\includegraphics[scale=1.2]{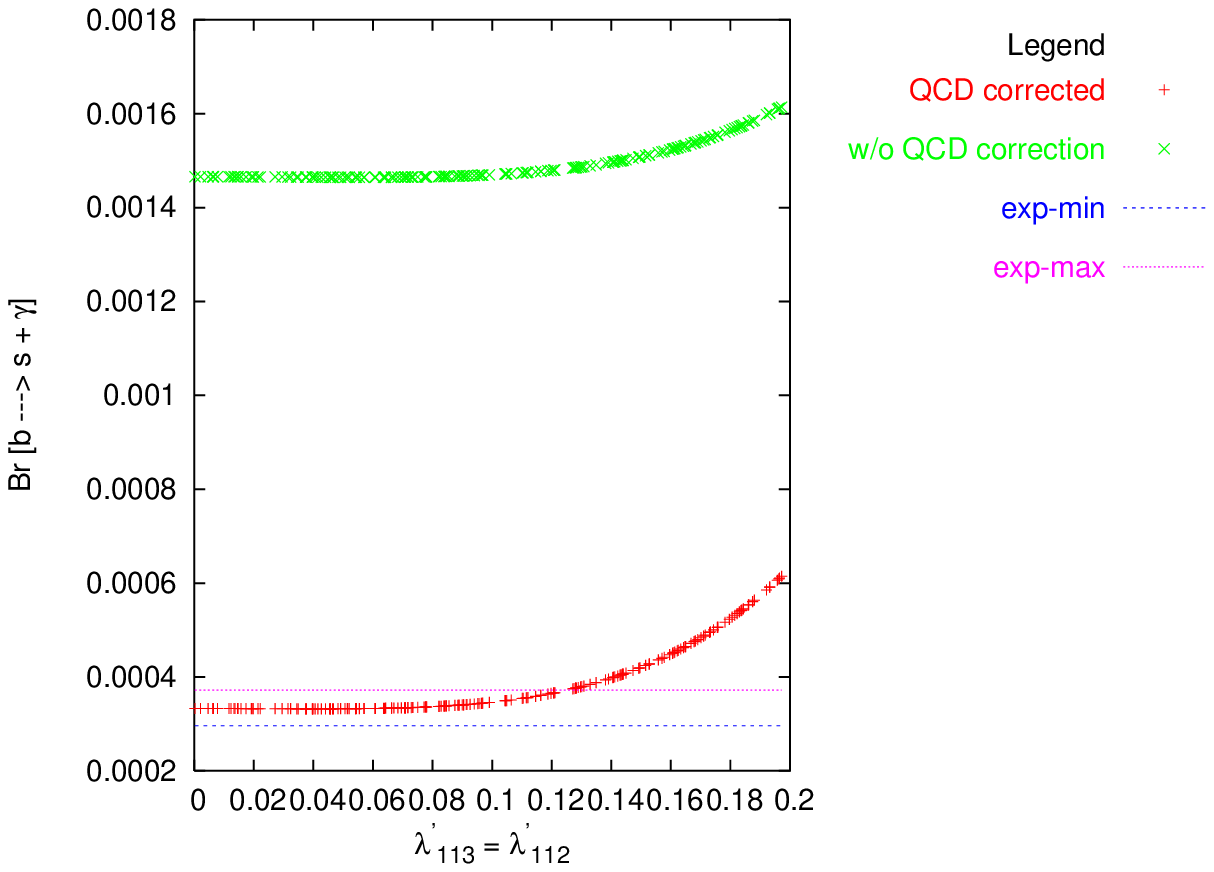}
\caption{\label{tt-113-112} Branching ratio as a function of
$\l'_{113} \;( =  \l'_{112})$. }
\end{figure}
The mild behaviour of branching fraction suggests the existence of cancellations. 
In Fig.\ref{tt-113-112} we have plotted branching fraction against 
$\l'_{113} \;( = \l'_{112})$ and obtain a bound of 
$|\l'_{113}\l'_{112}| < 1.4\times 10^{-2}$. Although the branching fraction behaves
more or less similarly in both the above cases, the underlying dynamics is quite 
different. To better appreciate the differences, we have plotted the relevant Wilson 
coefficients in Fig.\ref{ttw-233-232} for the combination $|\l'_{233}\l'_{232}|$ 
and in Fig.\ref{ttw-113-112} for the combination $|\l'_{113}\l'_{112}|$. In 
Fig.\ref{ttw-233-232}, the horizontal line (circles) is the large $R$-parity conserving
contribution to $C_7 (m_b)$ which does not interfere with the RPV
contributions going into $\wtl{C}_7$. The surprising feature is the large (though less
than $R$-parity conserving contribution) positive contribution of $\wtl{C}_7 (M_{\!\ssc W})$. 
This is surprising because it receives dominant negative sign contribution from the 
charged-slepton (crosses in the figure) whereas the positive sign sneutrino contribution 
(star marks in figure) is sub-dominant. Here again, the requirement of 
scheme-independence plays an important role because 
$\wtl{C}^{\rm eff}_7 (\m)=\wtl{C}_7 (\m) - \wtl{C}_{11} (\m)$. It is the large 
negative sign contribution from $\wtl{C}_{11} (M _{\!\ssc W})$ which adds up 
with the positive sneutrino contribution and dictates the behaviour of 
$\wtl{C}_7 (M _{\!\ssc W})$. At the scale $m_b$ after QCD running, there is some
suppression in the Wilson coefficient. Consider the case of the combination 
$|\l'_{113}\l'_{112}|$ (see Fig.\ref{ttw-113-112}). It has a peculiarity of the
charged-scalar loop contribution slightly dominating the contribution from
current-current Wilson coefficient $\wtl{C}_9$. This suggests that it is not 
always true that tree-level contribution dominates. Such a dominance can be 
traced to two effects : (a) In contrast to the tree level contribution, a charged slepton 
loop contribution has contributions from two diagrams corresponding to photon being 
emitted from a charged slepton ($\propto$ to loop function $F_2$) and
the photon being emitted from an up-type quark ($\propto F_1$). (b) The values of 
loop functions depend on the mass of the quark in the loop -- the lighter the quark 
less the suppression. Thus, provided that there is no CKM suppression, 
up-quark-charged-slepton loop dominates over the top-quark-charged-slepton loop 
and in fact can also dominate over the tree level contribution. That
there is no CKM suppression in both the above cases can be easily verified from the 
formulae of the Wilson coefficients. Also the uniform dominance of charged 
slepton loop over sneutrino loop can be accounted for by the argument (a) above. 
Since there are two possible diagrams for charged slepton loop it brings in a 
charge factor of (2/3 + 1 = 5/3) which is five times bigger than the charge
factor coming from the down-squark in the sneutrino loop. However, the 
charged-slepton and sneutrino loop contributions always come with opposite sign 
and hence there are cancellations.
\begin{figure}
\includegraphics[scale=1.2]{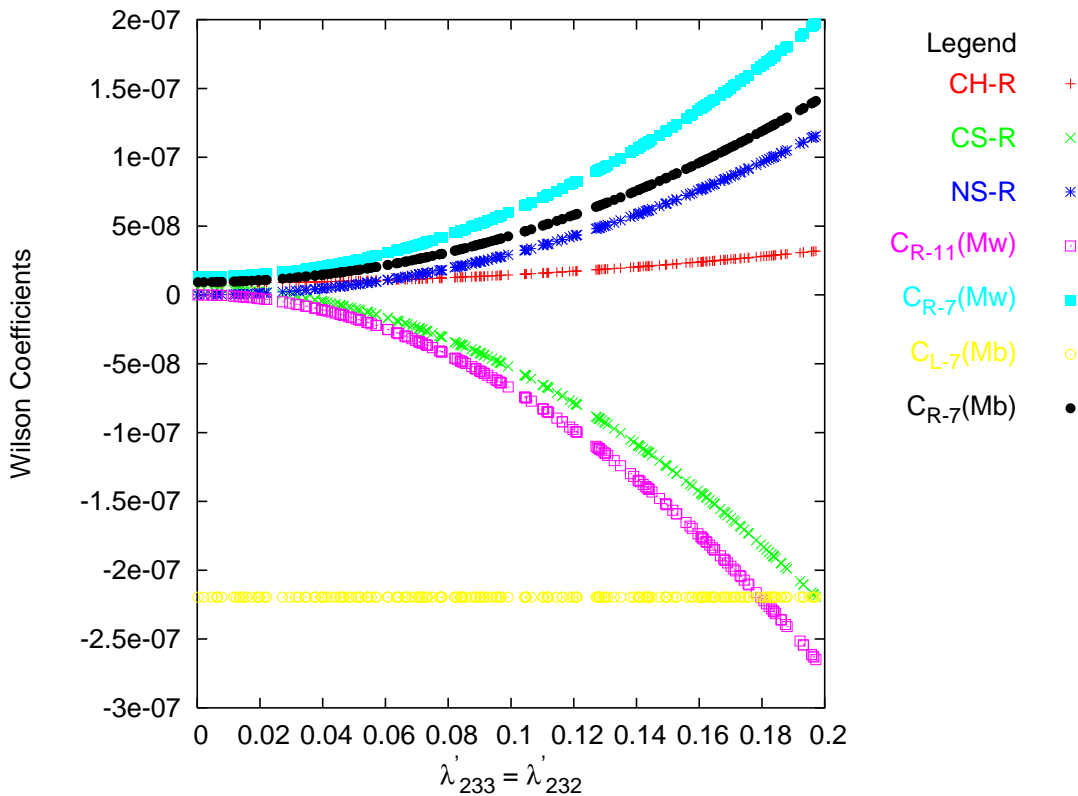}
\caption{\label{ttw-233-232}Relevant Wilson coefficients as a function
of $\l'_{233} \;(=\l'_{232})$.}
\end{figure}
\begin{figure}
\includegraphics[scale=1.2]{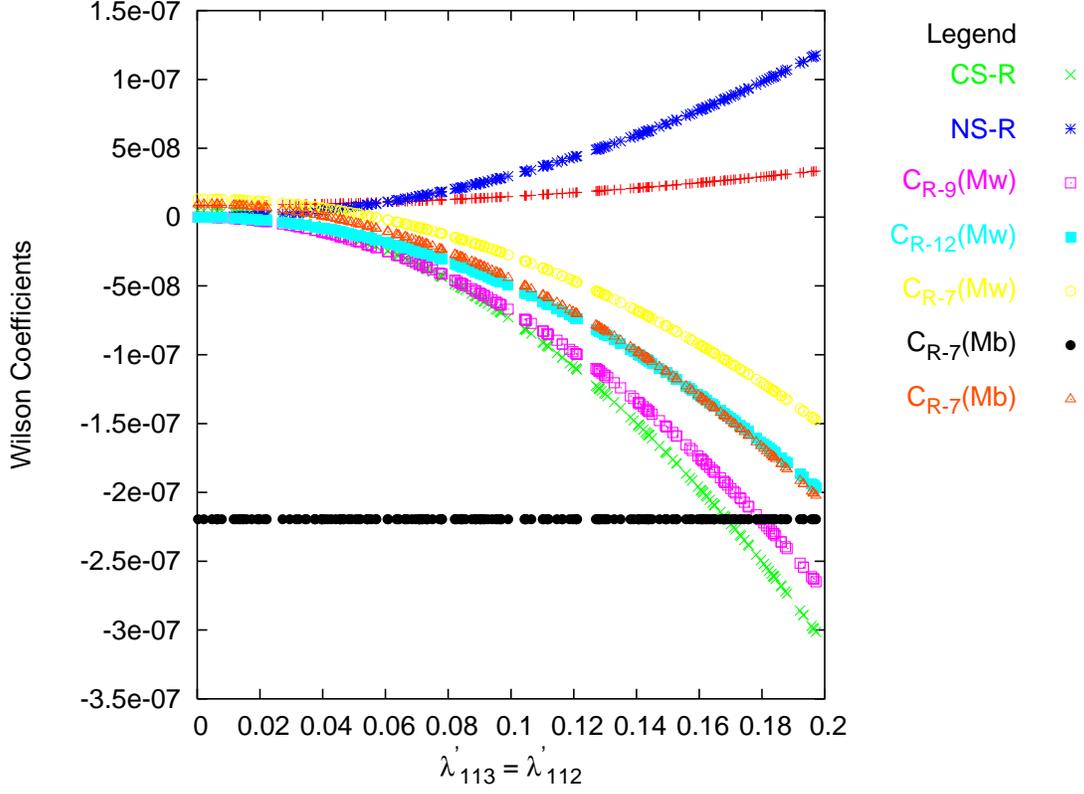}
\caption{\label{ttw-113-112}Relevant Wilson coefficients as a function
of $\l'_{113}\;(=\l'_{112})$.}
\end{figure}
\begin{table}
\caption{ Bounds on the various products of $\l'$ obtained
compared with the appropriately rescaled existing bounds \cite{gautam}}
\label{bounds}
\vspace{0.5cm}
\begin{tabular}{|l|l|l|l|}\hline
{\bf Product $|\l' \l'|$  } & {\bf Our bound} & {\bf Existing bound} & {\bf Wilson coeff.}\\\hline\hline
{\bf Case A} &&&\\\hline
131.121 &$4.9\times 10^{-3}$ &$2 \times 10^{-3}$& $C_7,C_9$ \\\hline
132.122 &$4.9\times 10^{-3}$& $3 \times 10^{-2}$  & $C_7,C_{10}$ \\\hline
133.123 &$1.6 \times 10^{-3}$ & $7.2 \times 10^{-5}$ & $C_7,C_{11}$ \\\hline
231.221 &$4.9\times 10^{-3}$&$2 \times 10^{-2}$  & $C_7,C_9$ \\\hline
232.222 &$4.9\times 10^{-3}$&$2 \times 10^{-2}$ & $C_7,C_{10}$\\\hline
233.223 &$1.6\times 10^{-3}$&$2 \times 10^{-2}$ & $C_7,C_{11}$\\\hline
331.321 &$4.9\times 10^{-3}$&$(1-3)\times 10^{-2}$ & $C_7,C_{9}$\\\hline
332.322 &$4.9\times 10^{-3}$&$(1-3)\times 10^{-2}$ & $C_7,C_{10}$\\\hline
333.323 &$1.6\times 10^{-3}$&$(1-3)\times 10^{-2}$ & $C_7,C_{11}$\\\hline\hline
{\bf Case B} &&&\\\hline
113.112 & $1.4 \times 10^{-2}$ & $1.3 \times 10^{-3}$& $\wtl C_7,\wtl C_9$\\\hline
123.122 & $1.4 \times 10^{-2}$& $1.3 \times 10^{-3}$ & $\wtl C_7, \wtl C_{10}$\\\hline
133.132 & $1.9 \times 10^{-2}$ & $7.2 \times 10^{-5}$ & $\wtl C_7, \wtl C_{11}$ \\\hline
213.212 & $1.4 \times 10^{-2}$ & $1.3 \times 10^{-3} $ & $\wtl C_7, \wtl C_9 $ \\\hline
223.222 & $1.4 \times 10^{-2}$ & $1.3 \times 10^{-3} $ & $ \wtl C_7, \wtl C_{10} $\\\hline
233.232 & $1.9 \times 10^{-2}$ & $ 4.4 \times 10^{-1}$ & $\wtl C_7, \wtl C_{11} $ \\\hline
313.312 & $ 1.4 \times 10^{-2} $ & $1.3 \times 10^{-3}$ & $\wtl C_7 , \wtl C_9$ \\\hline
323.322 & $ 1.4 \times 10^{-2} $ & $ 1.3 \times 10^{-3} $ & $\wtl C_7,\wtl C_{10}$ \\\hline 
333.332 &$1.9 \times 10^{-2}$ & $(1.3-9.2)\times 10^{-1}$ &$\wtl C_7,{\wtl C}_{11}$\\\hline
\end{tabular}
\end{table}

In Table\ref{bounds}, we have listed bounds that we have obtained for
the various $\l'$-coupling combinations among the cases (A) and (B). 
As far as one restricts oneself to the situation of two non-zero trilinear $\l'$-couplings,
case (A) and case (B)  are the only combinations that can contribute. With 
non-zero bilinears as well, or additional non-zero trilinears, there could
be more combinations that can contribute. For instance one can have a combination
of $\l'_{i3j}\l'_{hk2}$, giving rise to a neutrino or a neutral scalar loop 
for the penguins. Since the overall process conserves lepton number, the above
combination obviously needs to be supplemented with two more L-violating couplings.
This could be from a Majorana mass-insertions for neutrino or sneutrino in the loop. 
However, such a contribution would be further suppressed and hence we skip
presenting explicit results of the kind.

In all the above discussion, we have assumed that all the couplings are real and positive. 
It would be interesting
to consider the possibility when there is a relative phase resulting in a relative negative sign
among the chosen non-vanishing pair of $\l'$ couplings. It turns out that such a case does not 
affect the bounds much. For the case (B) with dominant contributions from RPV couplings, it obviously 
cannot affect the decay rate which is modulus squared of Wilson coefficient. However for case (A),
a negative or a positive sign can lead to constructive or destructive interference 
between $R$-parity conserving and violating contributions. A relative negative sign 
though changes the 
sign of the slope for the branching fraction it does not alter its magnitude. 
Thus, the only implication
as compared to the earlier bound which (for instance) was due to a branching fraction rising above 
the experimental limit, would now change to a
bound due to the branching fraction falling below the experimental limit. The change in the 
magnitude of the bound is however nominal.

Finally, we comment briefly on contributions from $\l^{\prime \prime}$-couplings.
We obtain no useful bounds on the  $\l^{\prime \prime}$-couplings. The main reason behind this is that
squark spectrum of 300 GeV is heavy enough to suppress the $\l^{\prime \prime}$ contributions.
However, a lighter spectrum (although not viable due to large MSSM contributions)
cannot lead to enhancement because of certain cancellations taking place. 
This provides a nice example of the interesting role played
by the QCD running and hence worth a brief discussion. 
The combinations responsible for contributing to the
decay rate can be written as: $\l^{\prime \prime *}_{hn3}\l^{\prime \prime}_{kn2}$ 
from the up-squark loop and
$\l^{\prime \prime *}_{nh3}\l^{\prime \prime}_{nk2}$ from the down squark loop, thus contributing to 
$\wtlc_{7,8}$. Depending on the particular
value for the indices, there could also be contributions from the current-current
operators with Wilson coefficients, $\wtlc_{1}( = -\wtlc_2),\wtlc_{14} 
(= -\wtlc_{15}),\wtlc_{16}(=-\wtlc_{17})$.
It might appear that the relative negative signs for the above coefficients leads to
cancellations and hence there is no impact on the decay rate. This is not what is
happening. There are cancellations, but these result rather from the form of 
$\wtlc_7$ at the scale $(m_b)$ (see eq.(\ref{c_mb}). 
Taking into account, the antisymmetry 
in the last two indices, one can form 18 combinations from the above mentioned general
form of the $\l''$ contributions (12 combinations contributing to the down-squark loop
and 9 to the up-squark loop but 3 three combinations common to both). There are two 
reasons that kill the contributions.  In a few cases, the loop level contribution
is accompanied by current-current Wilson coefficients pairs $\wtlc_{14}$ and
$\wtlc_{16}$, or $\wtlc_1$ and $\wtlc_{16}$. Although these pairs contribute about 
equal magnitude with identical sign, at the scale $m_b$ after QCD running these are
multiplied by about equal coefficients but with opposite sign and hence the 
cancellations.  The rest of the combinations get contributions from only one among 
the above pairs and hence cannot suffer from above cancellations. In this case, they all  
require flavor violating mass-insertions for (s)particles in the loop.
The only exception to above two cases is the combination 
$\l^{\prime \prime *}_{\ssc 313}\l^{\prime \prime}_{\ssc212}$ for which we get a bound of 0.5. 
This bound is very weak when compared to existing bound from other sources(
$6.2 \times 10^{-3}$\cite{gautam}).

\section{Conclusion}
In summary, we present a complete analysis of the decay rate \bsg\ at the leading log
order for the generic supersymmetric SM, or SUSY without $R$-parity. Unlike previous
studies on the topic, our results are fully generic, admitting all possible forms 
of $R$-parity violation without {\it a priori} assumption. We use exact mass eigenstates
in our formulae, hence free from the otherwise commonly used mass-insertion 
approximation. In case one prefers, our formulation does provide perturbative
approximations through which the explicit dependence on all RPV parameters can be 
extracted \cite{otto-gssm}. We consider the analytical results useful for any detailed
study on the model in relation to the radiative B decay.

In the numerical implementation of the analytical formulas our focus, is on the $\l'$ couplings.
For simplicity and feasibility, we keep two non-vanishing $\l'$-couplings.
The choice of the mass spectrum and various parameters of the model is dictated by the requirement 
that in the limit of  $R$-parity conserving SUSY, the branching ratio
falls within the experimental limits. Then, switching on the non-zero values of the
(combination of two) $\l'$-couplings allows us to trace their contribution and 
obtain bounds on the class of RPV parameters from imposing the experimental
constraints. In this aspect, the strategy is common to many of the earlier studies.
We find that RPV contributions not only expand the relevant four-quark operator
basis of the case of MSSM from 8 to 28, but also introduce new, likely 
to be dominating, contributions in the form of charged-slepton and sneutrino loops 
to the Wilson coefficients of (chromo)magnetic penguins. However, our results 
clearly show that the new four-quark operators with non-vanishing Wilson coefficients
at the electroweak/SUSY scale, induce effects through QCD running which 
typically dominate over the direct contributions to the Wilson coefficients of
(chromo)magnetic penguins. However, there are exceptions. Depending 
on the particles exchanged, the loop contributions could also dominate in some cases. 

The two-$\l'$ type of RPV contributions split into two classes: 
(A) $\l'_{i3j}\l^{\prime *}_{h2k}$ and (B) $\l^{\prime *}_{ij3}\l'_{hk2}$. 
Whereas the contributions in class (A) give rise to 
$b_{\!\ssc R} \rightarrow s_{\!\ssc L}$ transition and hence in direct interference 
with major contributions from SM and $R$-parity conserving SUSY parts, the 
contributions in class (B) have not much $R$-parity conserving counterparts. 
We obtain numerical bounds, some cases of which are orders of magnitude stronger 
than available bounds in the literature. The interpretation of such bounds has
to be taken more carefully. Their values depend strongly on the choice of
other SUSY parameters. Moreover, it could happen that the $R$-parity conserving
and RPV contributions have some accidental, but strong cancellation. In the latter 
case, no meaningful bounds on the RPV couplings alone can be obtained. After all,
the extra contributions in the case of MSSM are expected to have partial
cancellations to offset the strong charged Higgs contribution. Our numerical
results at least indicate clearly the strong implication of RPV couplings on
\bsg\ at a level substantially beyond earlier studies are able to illustrate. A most
conservative interpretation of the bounds obtained would be that they
indicate values of the $\l'$-couplings around and above which the corresponding
RPV contribution alone to  \bsg\  would be of alarming magnitude. We also  
explain clearly why the $\l''$-couplings have no major role to play in  \bsg .
Another type of significant contributions come from combinations of a bilinear
and a trilinear RPV couplings. The type of contributions has a quite different
structure and has not been studied before. They are implicitly included in our 
comprehensive formulation presented here, while the relevant numerical
study is to be presented in another publication\cite{017}. 

All the discussions above, like most of the earlier studies, has not taken into
consideration constraints on the RPV parameters from the neutrino masses. 
However, taking the reasonable assumption that all the neutrino masses are in 
the sub-eV range, and that there are no strong cancellations among RPV 
contributions, very stringent constraints on the parameters can be obtained 
\cite{kong-neutrino,kong-kang,rv-tri,cheung-kong}. In fact, such constraints are so important that talking about other numerical constraints on RPV parameters 
without reference to the neutrino mass constraints may not be a sensible way to 
take the model seriously. So, we would like to comment on the issue here. 
Majorana neutrino masses could be induced at one loop level by a pair of trilinear 
RPV couplings $\l'_{ilm}\l'_{jml}$. Because these violate lepton number by two units, 
they cannot contribute to the decay \bsg\ . However, the case $l=m$ could be used to
obtain bounds on $\l'_{imm}$.  Strong bounds on such individual couplings could
significantly strengthen the bounds on some of the combinations that contribute 
to \bsg\ . For instance, choosing the spectrum similar to the one for this study, 
one obtains a strong 
upper bound of $|\l'_{i33}| = 7.1 \times 10^{-4}$. The bound on the 
parameter $\l'_{i22}$ is much
less stringent ($|\l'_{i22}| = 3.0 \times 10^{-2}$) as it is 
suppressed by a quadratic factor of $(m_s/m_b)^2$ relative to the one-loop contribution
from the parameter
$\l'_{i33}$. This would mean that, whenever the coupling $\l'_{imm}$ is a part
of the combination contributing to \bsg\ , bounds coming from neutrino mass 
consideration are stronger. Nevertheless, the bounds obtained by us on three combinations
in case (B), namely $|\l'_{131}\l'_{121}|,|\l'_{231}\l'_{221}|,|\l'_{331}\l'_{321}|$,
still survive because $\l'_{imm}$ does not figure in these combinations.

\vspace{0.5cm}
\indent {\it note added in proof}: Just after this work was finished, an eprint 
\cite{neubert-new} appeared,
where the branching fraction for \Bsg\ has been recalculated at the NLL order 
using the results of soft-collinear factorization for inclusive $B$-meson decay
distributions. It has been pointed out that the singnificant perturbative uncertainties
associated with the parameter $\Delta = m_b - 2E_0$ have been ignored in the previous
works. The new
estimate now stands at $Br\left[B \rightarrow X_s + \g \;
(E_{\g} > 1.8 GeV) \right]_{\ssc \mathrm{SM}} = (3.44 \pm 0.53 \pm 0.35)\times 10^{-4}$,
where the first error is the estimate of perturbative uncertainties and the second one
reflects uncertainties in the input parameters. As a result of this larger theoretical
uncertainties, the lower bound on the charged-Higgs mass is strongly reduced compared to
previous estimates, to slightly below 200 GeV at 95 \% confidence level. With the larger
theoretical uncertainties in the SM prediction, the case for new physics  
certainly becomes stronger.

\vspace{0.5cm}

{\bf Acknowledgment:} 
O.K. thanks the Institute of Physics, Academia Sinica, Taiwan for 
hospitality during the early phase of the work.
R.V. acknowledges the hospitality of the Korea Institute for
Advanced Study (KIAS), Korea and KEK, Japan, during his visit where a part of the 
manuscript was written.
We also thank E.J.Chun, H.-N.Li, K. Hagiwara and K.S.Babu for the useful discussions 
and N.Oshimo for the
correspondence.
Our work is partially supported by research grants from the National
Science Council of Taiwan. O.K. is supported under grant number NSC
92-2112-M-008-044 and NSC 91-2112-M-008-042. R.V. is supported under
post-doc grant number NSC 92-2811-M-008-012 and NSC 92-2811-M-008-003.
\appendix
\section{The $\gamma$ matrix}
In  table \ref{table-L},\ref{tab-R} we provide a 
translation of our notation of operators to that of
refs.\cite{besmer,chun-bsg} for easy comparison. 
\label{gamma}
We define the scheme independent anomalous dimension matrix ${\hat \g}^{{\rm eff}}$ as:
\be
\label{gamma_eff}
{\hat \g}^{{(0)\rm eff}} =
\bmat{cc}
\g_{\!\ssc L}^{\rm eff} & 0\\
0& \g_{\!\ssc R}^{\rm eff}\\
\emat
\ee
where $\g_{\!\ssc L}^{\rm eff}$ 
and $\g_{\!\ssc R}^{\rm eff}$ are given as 
(for convenience we have given the operators on
the top row):
\be
\label{gamma_l}
\g_{\!\ssc L}^{\rm eff} = \ba{c}
\ba{rrrrrrrrrrr}
\cq_1&\;\cq_2&\;\cq_3&\;\cq_4&\;\cq_5&\;\cq_6&\;\;\cq_7&\;\;\cq_8&\;\,\cq_9&\;\,\cq_{10}&\;\,\cq_{11}
\ea \\
\left(
\ba{rrrrrrrrrrr}
-2&6&0&0&0&0&0&3&0&0&0\\
6&-2&-\f{2}{9}&\f{2}{3}&-\f{2}{9}&\f{2}{3}&\f{416}{81}&\f{70}{27}&
0&0&0\\
0&0&-\f{22}{9}&\f{22}{3}&-\f{4}{9}&\f{4}{3}&-\f{464}{81}&\f{545}{27}&
0&0&0\\
0&0&\f{44}{9}&\f{4}{3}&-\f{10}{9}&\f{10}{3}&\f{136}{81}&\f{512}{27}&
0&0&0\\
0&0&0&0&2&-6&\f{32}{9}&\f{-59}{3}&0&0&0\\
0&0&-\f{10}{9}&\f{10}{3}&-\f{10}{9}&-\f{38}{3}&-\f{296}{81}&-\f{703}{27}&
0&0&0\\
0&0&0&0&0&0&\f{32}{3}&0&0&0&0\\
0&0&0&0&0&0&-\f{32}{9}&\f{28}{3}&0&0&0\\
0&0&-\f{2}{9}&\f{2}{3}&-\f{2}{9}&\f{2}{3}&\f{200}{81}&-\f{119}{27}&
-16&0&0\\
0&0&-\f{2}{9}&\f{2}{3}&-\f{2}{9}&\f{2}{3}&\f{200}{81}&-\f{119}{27}&
0&-16&0\\
0&0&-\f{2}{9}&\f{2}{3}&-\f{2}{9}&\f{2}{3}&\f{200}{81}&-\f{227}{27}&0&
0&-16\\
\ea \right)
\ea
\ee
\linespread{1}
\bea &&
\g_{\!\ssc R}^{\rm eff} = 
\ba{c}
\ba{rrrrrrrrrrrrrrrrr}
\;\;\,{\cqt_1} & \, {\cqt_2} & \;\, {\cqt_3} &   {\cqt_4} & \;\,  {\cqt_{5}} 
& \;\; {\cqt_{6}}
& \,\,\;\; {\cqt_7} &  \,\;\;\, {\cqt_8} &  \; {\cqt_{9}} & \, {\cqt_{10}} & \,  {\cqt_{11}}
&  \, {\cqt_{12}} & \, {\cqt_{13}} &  {\cqt_{14}} &  \! {\cqt_{15}} & \! {\cqt_{16}} 
& \! {\cqt_{17}}
\ea \\
\left(
\ba{rrrrrrrrrrrrrrrrr}
-2&6&0&0&0&0&0&3&0&0&0&0&0&0&0&0&0\\
6&-2&-\f{2}{9}&\f{2}{3}&-\f{2}{9}&\f{2}{3}&\f{416}{81}&\f{70}{27}&0&0&0&0&0&0&0&0&0\\
0&0&-\f{22}{9}&\f{22}{3}&-\f{4}{9}&\f{4}{3}&-\f{464}{81}&\f{545}{27}&0&0&0&0&0&0&0&0&0\\
0&0&\f{44}{9}&\f{4}{3}&-\f{10}{9}&\f{10}{3}&\f{136}{81}&\f{512}{27}&0&0&0&0&0&0&0&0&0\\
0&0&0&0&2&-6&\f{39}{9}&-\f{59}{3}&0&0&0&0&0&0&0&0&0\\
0&0&-\f{10}{9}&\f{10}{3}&-\f{10}{9}&-\f{38}{3}&-\f{296}{81}&-\f{703}{27}&0&0&0&0&0&0&0&
0&0\\
0&0&0&0&0&0&\f{32}{3}&0&0&0&0&0&0&0&0&0&0\\
0&0&0&0&0&0&-\f{32}{9}&\f{28}{3}&0&0&0&0&0&0&0&0&0\\
0&0&-\f{2}{9}&\f{2}{3}&-\f{2}{9}&\f{2}{3}&\f{200}{81}&-\f{119}{27}&-16&0&0&0&0&0&0&0&0\\
0&0&-\f{2}{9}&\f{2}{3}&-\f{2}{9}&\f{2}{3}&\f{200}{81}&-\f{119}{27}&0&-16&0&0&0&0&0&0&0\\
0&0&-\f{2}{9}&\f{2}{3}&-\f{2}{9}&\f{2}{3}&\f{200}{81}&-\f{227}{27}&0&0&-16&0&0&0&0&0&0\\
0&0&-\f{2}{9}&\f{2}{3}&-\f{2}{9}&\f{2}{3}&-\f{448}{81}&-\f{119}{27}&0&0&0&-16&0&0&0&0&0\\
0&0&-\f{2}{9}&\f{2}{3}&-\f{2}{9}&\f{2}{3}&-\f{448}{81}&-\f{119}{27}&0&0&0&0&-16&0&0&0&0\\
0&0&0&0&0&0&0&3&0&0&0&0&0&-2&6&0&0\\
0&0&-\f{2}{9}&\f{2}{3}&-\f{2}{9}&\f{2}{3}&\f{416}{81}&\f{70}{27}&0&0&0&0&0&6&-2&0&0\\
0&0&0&0&0&0&0&3&0&0&0&0&0&0&0&-2&6\\
0&0&-\f{2}{9}&\f{2}{3}&-\f{2}{9}&\f{2}{3}&-\f{232}{81}&\f{70}{27}&0&0&0&0&0&0&0&6&-2\\
\ea \right)
\ea \nonumber \\
&& \label{gamma_r}
\eea
\begin{table}[h]
\caption{Operators that mix in the block $\g_{\!\ssc L}^{\rm eff}$ }
\label{table-L}
\begin{tabular}{|l|l|l|l|l|l|l|l|l|l|l|l|}\hline
{\bf Our notation}&$\cq_1$&$\cq_2$&$\cq_3$&$\cq_4$&$\cq_5$&$\cq_6$&$\cq_7$&$\cq_8$
&$\cq_9$&$\cq_{10}$&$\cq_{11}$\\\hline
{\bf Ref.\cite{chun-bsg}}&$O_{\!\ssc 1L}$&$O_{\!\ssc 2L}$&$O_{\!\ssc 3L}$&
$O_{\!\ssc 4L}$&$O_{\!\ssc 5L}$&$O_{\!\ssc 6L}$
&$O_{\!\ssc 7L}$&$O_{\!\ssc 8L}$&$O^d_{\!\ssc 6L}$&$O^s_{\!\ssc 6L}$
&$O^b_{\!\ssc 6L}$\\\hline
{\bf Ref.\cite{besmer}}&$O_1$&$O_2$&$O_3$&$O_4$&$O_5$&$O_6$&$O_7$&$O_8$&$P_{\ssc 6}$&
$P_{\ssc 7}$&$P_{\ssc 8}$\\\hline
\end{tabular}
\end{table}
\begin{table}[h]
\caption{Operators that mix in the block $\g_{\!\ssc R}^{\rm eff}$ .}
\label{tab-R}
\begin{tabular}{|l|l|l|l|l|l|l|l|l|l|l|l|l|l|l|l|l|l|}\hline
{\bf Our notation}&${\cqt_1}$ &$ {\cqt_2}$ &$ {\cqt_3}$ &$ {\cqt_4}$ &$ {\cqt_{5}}$ 
&$ {\cqt_{6}}$
&${\cqt_7}$ &$ {\cqt_8}$ &$ {\cqt_{9}}$ &$ {\cqt_{10}}$ &$ {\cqt_{11}}$
& ${\cqt_{12}}$ &$ {\cqt_{13}}$ &$ {\cqt_{14}}$ &$ {\cqt_{15}}$ &$ {\cqt_{16}} $
& ${\cqt_{17}}$\\\hline
{\bf Ref.\cite{chun-bsg}}&$O_{\!\ssc 1R}$&$O_{\!\ssc 2R}$&$O_{\!\ssc 3R}$&$O_{\!\ssc 4R}$&$O_{\!\ssc 5R}$&$O_{\!\ssc 6R}$&
$O_{\!\ssc 7R}$&$O_{\!\ssc 8R}$&$O^d_{\!\ssc 6R}$&$O^s_{\!\ssc 6R}$&$O^b_{\!\ssc 6R}$&$O^u_{\!\ssc 6R}$&$O^c_{\!\ssc 6R}$
&$O^u_{\!\ssc 3R}$&$O^u_{\!\ssc 4R}$&$O^d_{\!\ssc 3R}$&$O^d_{\!\ssc 4R}$\\\hline
{\bf Ref.\cite{besmer}}&$R_{\ssc 3}$&$R_{\!\ssc 4}$&$P_{\ssc 9}$
&$P_{\ssc 10}$&$P_{\ssc 11}$&$P_{\ssc 12}$&${\wtl O}_7$&
${\wtl O}_8$&$P_{\ssc 3}$&$P_{\ssc 4}$&$P_{\ssc 5}$&$P_{\ssc 1}$&$P_{\ssc 2}$&$R_{\ssc 1}$&$R_{\ssc 2}$&$R_{\ssc 5}$&$R_{\ssc 6}$\\\hline
\end{tabular}
\end{table}



\begin{thebibliography}{99}
\bibitem{cella}Scheme-independence was further confirmed using an unconventional scheme
by G. Cella \atal\ \np{B421}{417}{94},\pl{B325}{227}{94}.
\end{thebibliography}
\end{document}